\documentclass[sigconf]{aamas/aamas}
\usepackage{balance}

\usepackage[utf8]{inputenc} %
\usepackage[T1]{fontenc}    %
\usepackage{hyperref}       %
\usepackage{url}            %

\usepackage{graphicx}
\usepackage{amsmath, amsthm, amsfonts, mathtools, physics, bm, bbm, optidef, nicefrac, xfrac}
\usepackage{float, subcaption, wrapfig}
\usepackage{csquotes}
\usepackage{multirow, booktabs}
\usepackage{algorithm, algorithmic}
\usepackage[capitalise,noabbrev]{cleveref}
\usepackage{parskip, microtype}
\usepackage[inline]{enumitem}
\usepackage{siunitx}
\usepackage{comment}
\usepackage{tikz, pgfplots, pgfplotstable}
\pgfplotsset{compat=1.18}
\usetikzlibrary{shapes,arrows,automata,positioning}
\usepackage{xcolor}         %
\definecolor{InitialColor}{RGB}{255, 100, 0}
\definecolor{TrialColor}{RGB}{40, 40, 255}
\definecolor{EvalColor}{RGB}{0, 128, 10}

\usepackage{acro}
\DeclareAcronym{MCTS}{%
	short=MCTS,%
	short-plural=s,%
	short-indefinite=an,%
	long=Monte-Carlo tree search,%
	long-plural=s,%
	long-indefinite=a}

\DeclareAcronym{AHT}{%
	short=AHT,%
	short-indefinite=an,%
	long=ad hoc teamwork,%
	long-indefinite=a}

\DeclareAcronym{PPO}{%
	short=PPO,%
	short-indefinite=a,%
	long=proximal policy optimisation,%
	long-indefinite=a}
 
\DeclareAcronym{IPPO}{%
	short=IPPO,%
	short-indefinite=an,%
	long=independent PPO,%
	long-indefinite=an}
 
\DeclareAcronym{MAPPO}{%
	short=MAPPO,%
	short-indefinite=an,%
	long=multi-agent PPO,%
	long-indefinite=a}

\DeclareAcronym{A2C}{%
	short=A2C,%
	short-indefinite=a,%
	long=advantage actor-critic,%
	long-indefinite=a}

\DeclareAcronym{RL}{%
	short=RL,%
	short-indefinite=an,%
	long=reinforcement learning,%
	long-indefinite=a}

\DeclareAcronym{MARL}{%
	short=MARL,%
	short-indefinite=a,%
	long=multi-agent reinforcement learning,%
	long-indefinite=a}

\DeclareAcronym{MDP}{%
	short=MDP,%
	short-plural=s,%
	short-indefinite=an,%
	long=Markov decision process,%
	long-plural=es,%
	long-indefinite=a}

\DeclareAcronym{POMDP}{%
	short=POMDP,%
	short-plural=s,%
	short-indefinite=a,%
	long=partially observable Markov decision process,%
	long-plural=es,%
	long-indefinite=a}

\DeclareAcronym{Dec-POMDP}{%
	short=Dec-POMDP,%
	short-plural=s,%
	short-indefinite=a,%
	long=decentralised partially observable Markov decision process,%
	long-plural=es,%
	long-indefinite=a}

\DeclareAcronym{Dec-MDP}{%
	short=Dec-MDP,%
	short-plural=s,%
	short-indefinite=a,%
	long=decentralised Markov decision process,%
	long-plural=es,%
	long-indefinite=a}

\DeclareAcronym{SG}{%
	short=SG,%
	short-plural=s,%
	short-indefinite=an,%
	long=stochastic game,%
	long-plural=s,%
	long-indefinite=a}

\DeclareAcronym{POSG}{%
	short=POSG,%
	short-plural=s,%
	short-indefinite=a,%
	long=partially observable stochastic game,%
	long-plural=s,%
	long-indefinite=a}

\DeclareAcronym{MAS}{%
	short=MAS,%
	short-plural=s,%
	short-indefinite=an,%
	long=multi-agent system,%
	long-plural=s,%
	long-indefinite=a}

\DeclareAcronym{SAS}{%
	short=SAS,%
	short-plural=s,%
	short-indefinite=an,%
	long=single-agent system,%
	long-plural=s,%
	long-indefinite=a}

\DeclareAcronym{AI}{%
	short=AI,%
	short-indefinite=an,%
	long=artificial intelligence,%
	long-plural=s,%
	long-indefinite=an}

\DeclareAcronym{MLE}{%
	short=MLE,%
	short-plural=s,%
	short-indefinite=an,%
	long=maximum likelihood estimation,%
	long-plural=s,%
	long-indefinite=a}

\DeclareAcronym{MAP}{%
	short=MAP,%
	short-plural=s,%
	short-indefinite=a,%
	long=maximum a posteriori,%
	long-plural=s,%
	long-indefinite=a}

\DeclareAcronym{IID}{%
	short=IID,%
	short-plural=s,%
	short-indefinite=an,%
	long=identically and independently distributed,%
	long-plural=s,%
	long-indefinite=an}

\DeclareAcronym{PMF}{%
	short=PMF,%
	short-plural=s,%
	short-indefinite=a,%
	long=probability mass function,%
	long-plural=s,%
	long-indefinite=a}

\DeclareAcronym{PDF}{%
	short=PDF,%
	short-plural=s,%
	short-indefinite=a,%
	long=probability density function,%
	long-plural=s,%
	long-indefinite=a}

\DeclareAcronym{CDF}{%
	short=CDF,%
	short-plural=s,%
	short-indefinite=a,%
	long=cumulative density function,%
	long-plural=s,%
	long-indefinite=a}

\DeclareAcronym{MEDoE}{%
	short=MEDoE,%
	short-plural=s,%
	short-indefinite=a,%
	long=Modulating Exploration and Training via Domain of Expertise,%
	long-plural=s,%
	long-indefinite=a}

\DeclareAcronym{DoE}{%
	short=DoE,%
	short-plural=s,%
	short-indefinite=a,%
	long=domain of expertise,%
	long-plural-form=domains of expertise,%
	long-indefinite=a}

\DeclareAcronym{FST}{%
	short=FST,%
	short-plural=s,%
	short-indefinite=a,%
	long=Few-Shot Teamwork,%
	long-plural=s,%
	long-indefinite=a}

\DeclareAcronym{MLP}{%
	short=MLP,%
	short-plural=s,%
	short-indefinite=an,%
	long=multi-layer perceptron,%
	long-plural=s,%
	long-indefinite=a}
 
\DeclareAcronym{KL}{%
	short=KL,%
	short-plural=s,%
	short-indefinite=a,%
	long=Kullback-Leibler,%
	long-plural=s,%
	long-indefinite=a}
 
\DeclareAcronym{AUC}{%
	short=AUC,%
	short-plural=s,%
	short-indefinite=an,%
	long=area under curve,%
	long-plural=s,%
	long-indefinite=an}

\DeclareAcronym{CTDE}{%
	short=CTDE,%
	short-plural=s,%
	short-indefinite=a,%
	long=centralised training decentralised execution,%
	long-plural=s,%
	long-indefinite=a}

\DeclareAcronym{STD}{%
	short=STC,%
	short-plural=s,%
	short-indefinite=an,%
	long=sub-task curriculum,%
	long-plural-form=sub-task curricula,%
	long-indefinite=a}

\DeclareMathOperator{\Loss}{\mathcal{L}}

\DeclareMathOperator{\Expect}{\mathbb{E}}

\DeclarePairedDelimiterX{\infdivx}[2]{(}{)}{#1\;\delimsize\|\;#2}
\newcommand{\KL}{D_{KL}\infdivx}

\newcommand{\iAgent}{i}

\newcommand{\AgentSet}{A}

\newcommand{\State}{s}
\newcommand{\StateSpace}{\mathcal{S}}

\newcommand{\Obs}{o}
\newcommand{\ObsFunc}{O}
\newcommand{\ObsSpace}{\Omega}

\newcommand{\Action}{a}

\newcommand{\ActionSpace}{\mathcal{A}}
\newcommand{\StateTransFunc}{T}

\newcommand{\Reward}{r}
\newcommand{\RewardFunc}{R}

\newcommand{\Policy}{\pi}
\newcommand{\PolicySpace}{\Pi}
\newcommand{\PolicyParams}{\theta}
\newcommand{\VParams}{\psi}
\newcommand{\Discount}{\gamma}

\newcommand{\Task}{\mathcal{T}}
\newcommand{\Curric}{\mathcal{C}}

\newcommand{\iTarget}{T}

\newcommand{\Team}{I}
\newcommand{\LearnAlg}{\mathbb{L}}

\newcommand{\DoEC}{D}
\newcommand{\DoECL}{\hat{\DoEC}}
\newcommand{\DoESet}{\mathcal{E}}
\newcommand{\joint}{\vb}

\newcommand{\Temp}{T}
\newcommand{\EntCoef}{\alpha}
\newcommand{\KLCoef}{\kappa}

\newcommand{\Base}{\text{base}}
\newcommand{\BoostCoef}{\beta}
\newcommand{\TempBoostCoef}{\BoostCoef_\Temp}
\newcommand{\EntBoostCoef}{\BoostCoef_\EntCoef}
\newcommand{\KLBoostCoef}{\BoostCoef_\KLCoef}

\newcommand{\Tmax}{T_\text{max}}
\newcommand{\nStep}{n}
\newcommand{\ImpWeight}{w}
\newcommand{\InitialState}{\mu}

\newcommand\blfootnote[1]{%
\begingroup
\renewcommand\thefootnote{}\footnote{#1}%
\addtocounter{footnote}{-1}%
\endgroup
}

\setcopyright{ifaamas}
\acmConference[AAMAS '24]{Proc.\@ of the 23rd International Conference
on Autonomous Agents and Multiagent Systems (AAMAS 2024)}{May 6 -- 10, 2024}
{Auckland, New Zealand}{N.~Alechina, V.~Dignum, M.~Dastani, J.S.~Sichman (eds.)}
\copyrightyear{2024}
\acmYear{2024}
\acmDOI{}
\acmPrice{}
\acmISBN{}

\acmSubmissionID{782}

\title{Learning Complex Teamwork Tasks Using a Given Sub-task Decomposition}

\author{Elliot Fosong}
\affiliation{
  \institution{University of Edinburgh}
  \city{Edinburgh}
  \country{United Kingdom}}
\email{e.fosong@ed.ac.uk}

\author{Arrasy Rahman}
\affiliation{
  \institution{University of Texas at Austin}
  \city{Austin, TX}
  \country{USA}}
\email{arrasy@cs.utexas.edu}

\author{Ignacio Carlucho}
\affiliation{
  \institution{Heriot-Watt University}
  \city{Edinburgh}
  \country{United Kingdom}}
\email{ignacio.carlucho@hw.ac.uk}

\author{Stefano V. Albrecht}
\affiliation{
  \institution{University of Edinburgh}
  \city{Edinburgh}
  \country{United Kingdom}}
\email{s.albrecht@ed.ac.uk}

\begin{abstract}
	Training a team to complete a complex task via \ac{MARL} can be difficult due to challenges such as policy search in a large joint policy space, and non-stationarity caused by mutually adapting agents.  
  To facilitate efficient learning of complex multi-agent tasks, we propose an approach which uses an expert-provided decomposition of a task into simpler multi-agent sub-tasks.  
  In each sub-task, a subset of the entire team is trained to acquire sub-task-specific policies. The sub-teams are then merged and transferred to the target task, where their policies are collectively fine-tuned to solve the more complex target task.   
  We show empirically that such approaches can greatly reduce the number of timesteps required to solve a complex target task relative to training from-scratch.
  However, we also identify and investigate two problems with naive implementations of approaches based on sub-task decomposition, and propose a simple and scalable method to address these problems which augments existing actor-critic algorithms.
  We demonstrate the empirical benefits of our proposed method, enabling sub-task decomposition approaches to be deployed in diverse multi-agent tasks.
\end{abstract}


\makeatletter

\begin{document}

\pagestyle{fancy}
\fancyhead{}

\maketitle 

\section{Introduction}\label{sec:introduction}
    In cooperative \acf{MARL} \citep{albrechtMultiagentReinforcementLearning2024}, the goal is to have a team of autonomous agents learn to complete a task, by having the team gather and learn from experiences in that task.
    Although \ac{MARL} techniques have been used successfully to solve a range of cooperative team-based tasks, there are still challenges in complex scenarios.
	These challenges include multi-agent credit assignment \cite{changAllLearningLocal2003}, non-stationarity due to simultaneously adapting agents \cite{papoudakisDealingNonstationarityMultiagent2019}, difficulty searching over a large joint action space, and equilibrium selection problems \citep{clausDynamicsReinforcementLearning1998,weiLenientLearningIndependentlearner2016}.
    These problems typically worsen when the number of agents increases, or when complex coordination is required.

    We propose addressing these problems and solving complex multi-agent tasks by using a curriculum of sub-tasks.
    We focus on complex cooperative tasks that may be decomposed into sub-tasks, where each sub-task could be solved by a subset of the agents. 
    We start by training sub-teams of agents on their respective sub-tasks before fine-tuning the full set of agents on the target task.
    In doing so, we induce a curriculum for the agents, allowing them to bypass the initial stages of random search by leveraging skills acquired during the first phase of training.
	Furthermore, by training initially in simpler tasks with fewer agents, we reduce the problems caused by non-stationary and multi-agent credit assignment.
    While we assume this sub-task decomposition is given, many tasks have a natural decomposition into sub-tasks that could use our proposed methodology.
	\blfootnote{Code and experimental data available at \url{https://github.com/uoe-agents/MEDoE}}

\begin{figure}[t]
	\centering
    \includegraphics[width=\linewidth]{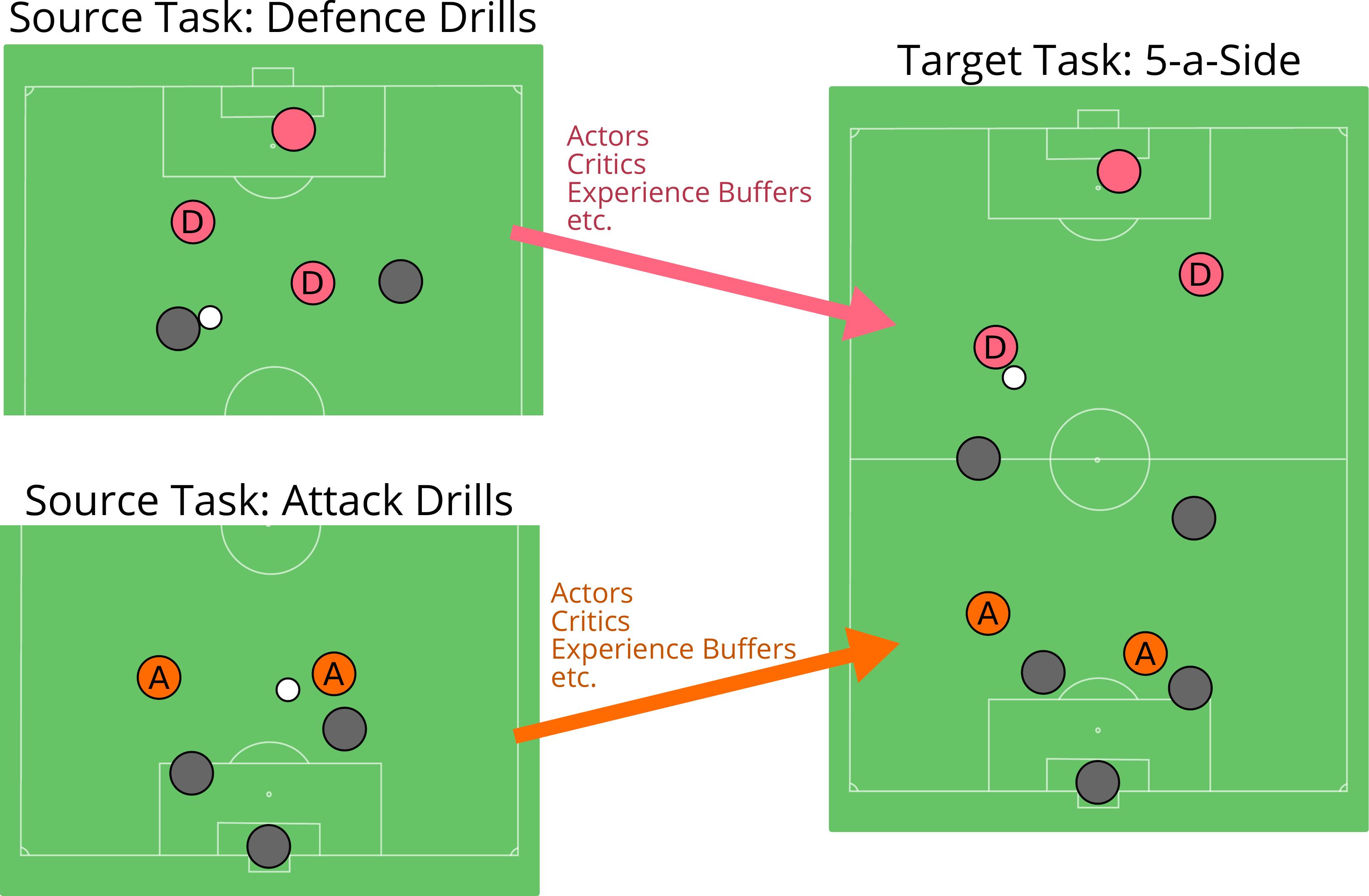}
	\caption{
		\Acl{STD} for 5-a-side football. We control the red team, and play against the grey team.
		We train defender agents in defensive drills (top-left) and attacker agents in attack drills (bottom-left).
		As represented by the arrows, we transfer these agents to the 5-a-side target task.
		We then fine-tune our combined team in the target task.
	}
	\label{fig:curric}
\end{figure}
    
    For example, consider training five agents to play a 5-a-side football target task by breaking the problem up into two sub-tasks: attack drills with two attackers, and defence drills with two defenders and one goalkeeper (\cref{fig:curric}).
    The attackers learn skills including \enquote{shooting on target} and \enquote{avoiding being tackled} which are useful in the full 5-a-side football game.  
	Likewise, defenders learn skills useful in 5-a-side football such as \enquote{blocking} and \enquote{tackling}.
    When the attackers and defenders are recombined, extra fine-tuning is required in the full 5-a-side football game, for example, to teach defenders that they ought to pass to their attacker teammates.

	We conjecture that \ac{MARL} approaches can learn complex target tasks more efficiently by first training policies in each sub-task, followed by combining the sub-task policies and fine-tuning them in the target task.
	We explore the feasibility of this approach in experiments with human-designed sub-task curricula using standard \ac{MARL} techniques building on top of the \ac{PPO} algorithm \cite{schulmanProximalPolicyOptimization2017}.
	We find that using such an approach, we can learn to solve multi-agent tasks which are difficult for current state-of-the-art \ac{MARL} algorithms, suggesting an interesting avenue for future \ac{MARL} research and practitioners.
	
	However, we also identify two issues with the naive application of standard \ac{MARL} techniques to fine-tuning:
	\begin{enumerate*}
		\item miscoordinated exploration leading to convergence to sub-optimal equilibria; and
		\item agents forgetting useful skills they obtained during sub-task training.
	\end{enumerate*}
	To address these issues, we further propose an approach called \emph{\ac{MEDoE}} which automatically infers based on sub-task experience replay buffers the circumstances in which an agent has expertise in the complex target task; and then uses the expertise predictions to modulate hyperparameters in the fine-tuning process.
	\Ac{MEDoE} is a flexible approach that can extend any decentralised execution actor-critic method, and can be used with any number of agents.
    Furthermore, \ac{MEDoE} can also be used in situations where the size of the team in the complex target task may not be known during training in the simple sub-tasks, allowing for flexibility with respect to the target team composition, and re-use of pre-trained agents in fine-tuning for different downstream target tasks.  

	Continuing our football example, \ac{MEDoE} might predict that in a defensive scenario in 5-a-side football (\cref{fig:curric}, right) the defender agents have expertise --- i.e., their policies are already near-optimal.
	Assuming these defenders do indeed have expertise, \ac{MEDoE} reduces the likelihood the defenders take exploratory actions, reducing variance caused by exploration, thereby stabilising training of other agent policies.
	Additionally, \ac{MEDoE} reduces the rate at which the defenders forget their existing skills by increasing the coefficient of penalty terms in the policy loss for deviating from their final sub-task policies, and decreasing the entropy bonus coefficient.

	Our experiments show that sub-task decomposition methods can solve complex teamwork tasks in many fewer training timesteps than baseline methods which train from scratch in the target task. In some environments, naive fine-tuning approaches are sufficient to outperform from-scratch baselines, but in others our proposed method \ac{MEDoE} is responsible for the improved performance of using sub-task decompositions.

\section{Problem Formulation}\label{sec:formulation}
In this section, we define the framework for our sub-task decomposition approach to accelerating \ac{MARL}.

\subsection{Sub-task Decomposition}\label{sec:curriculum}
We describe our sub-task decomposition approach in terms of \emph{\acp{STD}}, which model how a sub-task decomposition can be used in training.
In this work, we consider cases where we are provided with \iacl{STD}, and distinguish our work from those which seek to learn the curriculum.
\Iacl{STD} $\Curric$ is a tuple,
$\langle \Task, \qty{\Curric_\text{sub.}}, \AgentSet, \LearnAlg \rangle$.
The task $\Task$ is a \acs{Dec-POMDP}\cite{oliehoekConciseIntroductionDecentralized2016},
\begin{equation}
	\mathcal{M} = 
        \langle
            \Team,
            \StateSpace,
            \qty{\ActionSpace_i}_{i\in\Team}, 
            \StateTransFunc,
			\InitialState,
            \qty{\ObsSpace_i}_{i\in\Team}, 
            \ObsFunc,
            \RewardFunc,
            \gamma
        \rangle,
\end{equation}
where
$\Team$ is the set of agents (the team);
$\StateSpace$ is the state space;
$\ActionSpace_\iAgent$ is the action space for agent $\iAgent$;
$\StateTransFunc(\State_{t+1}|\State_t,\Action_t)$ is the state transition probability density function;
$\InitialState(\State_0)$ is the initial state distribution;
$\ObsSpace_\iAgent$ is the observation space for agent $\iAgent$;
$\ObsFunc(\Obs_t|\State_t,\Action_{t-1})$ is the observation probability density function;
$\RewardFunc: \StateSpace\times\ActionSpace\mapsto\mathbb{R}$ is the team reward function;
and
$\gamma$ is the discount factor.
The objective within \iacs{Dec-POMDP} is to find a joint policy $\pi$ which maximises the expected discounted return $G=\Expect_{}\bqty{\sum_{t}\gamma^t\RewardFunc(\State_t,\Action_t)}$.

$\qty{\Curric_\text{sub.}}$ is a set of \aclp{STD}, where each \acl{STD} is defined recursively. 
The base-case for recursion, $\qty{\Curric_\text{sub.}} = \emptyset$, corresponds to training from scratch on task $\Task$.
We also refer to sub-tasks as \emph{source tasks}, using transfer learning terminology \cite{silvaSurveyTransferLearning2019}.

$\AgentSet$ is a function which maps from sub-tasks $\qty{\Curric_\text{sub.}}$ to a set of agents $\Team$ for task $\Task$. Each agent has associated data --- in this paper we use actor-critic methods, so each agent has an associated policy $\pi$, value function $V$, and we also include an experience replay buffer for each agent. For conciseness we also use $\AgentSet$ to represent the set of agents which initialise the current task in the curriculum.

$\LearnAlg$ is a learning algorithm applied in each task of the \acl{STD}, which given the agents $\AgentSet$ and \acs{Dec-POMDP} $\Task$ learns policies for task $\Task$. For example, in this paper $\LearnAlg$ is the \acs{IPPO} algorithm with particular hyperparameters and stopping conditions.

Though we present a general framework for modelling sub-task curricula, in this work, we focus on curricula with tree-depth of one.
As a concrete example, the 5-a-side football task from \cref{fig:curric} the \acl{STD} would be:
\begin{itemize}
	\item $\Curric = \langle \Task_\text{5v5}, \qty{\Curric_\text{def.}, \AgentSet,  \Curric_\text{att.}} , \LearnAlg \rangle$
	\item $\Curric_\text{def} = \langle \Task_\text{def.}, \emptyset , \AgentSet , \LearnAlg \rangle$
	, $\Curric_\text{att.} = \langle \Task_\text{att.}, \emptyset , \AgentSet , \LearnAlg \rangle$
\end{itemize}

\subsection{Objective}\label{sec:objective}
Our objective is to accelerate the rate at which \ac{MARL} learns to solve a complex target task.
Here we define our objective in terms of minimising the total number of timesteps taken to reach a desired level of performance in the target task.

Let $N(\Curric)$ be the number of training timesteps in the current task of the curriculum $\Curric$, and $N^\text{tot.}(\Curric)$ be the \emph{total} number of training timesteps used in the entire \acl{STD} $\Curric_i$, i.e., 
\begin{equation*}
	N^\text{tot.}(\Curric) = N(\Curric) + \sum_{\Curric_s \in \qty{\Curric_\text{sub.}}}{N^\text{tot.}(\Curric_s)}
\end{equation*}

Given a target task $\Task_\text{target}$ and target performance $G_\text{target}$ our overall objective is to find a sub-task curriculum $\Curric^*_\text{target} = \langle \Task_\text{target}, \cdot, \cdot, \cdot \rangle$ which minimises $N(\Curric^*_\text{target})$ subject to the constraint that the produced agents attain expected returns $G \geq G_\text{target}$.

This optimisation problem has many free variables including the learning algorithm used at each point in the curriculum; the stopping conditions at each point in the curriculum; the sub-tasks used at each point in the curriculum; and the agents transferred at each point in the curriculum.
In this work, we first investigate the feasibility of the approach based on standard \ac{MARL} components without attempting to optimise the sub-task curriculum.
We then propose an approach to improve performance by modifying the learning algorithm $\LearnAlg_\text{target}$ used in the target task.

\section{Sub-task Curricula with Standard MARL Approaches}\label{sec:naive-stc}
In this section, we investigate applying standard \ac{MARL} techniques for fine-tuning in the target tasks of sub-task curricula. We find that while this naive approach can sometimes work, we identify and analyse two problems which can arise.
In \cref{sec:medoe} we present a method which addresses these problems and greatly improves the performance of sub-task curriculum approaches in some tasks.

\subsection{Environments}\label{sec:environments}
To test the sub-task curriculum approach, we consider three environments with clear task decompositions:
\emph{Chainball}, a simple but difficult to solve environment we introduce to provide insight; %
\emph{Overcooked} \citep{wangTooManyCooks2020,rotherDisentanglingInteractionUsing2023}, a gridworld environment common in \ac{MARL} research; %
and \emph{VMAS Football} \citep{bettiniVMASVectorizedMultiagent2022}, a complex 2D physics-based football simulation.
These tasks are fully-observable,
and we ensure that observation dimensions are consistent between tasks within the curriculum by zero-padding observations where necessary.
Further details of each environment can be found in \cref{sec:environment-details}.

\begin{figure}[t]
	\begin{subfigure}[t]{\linewidth} %
		\centering
		\scalebox{0.78}{
			\begin{tikzpicture}%
				[>=stealth,
				 shorten >=0pt,
				 node distance=1cm,
				 on grid,
				 auto,
				]
				\node[state]	(mid)						{M};
				\node[state]	(def)	[left = of mid]		{D};
				\node[state]	(att)	[right = of mid]	{A};
				\node[state]	(gkc)	[left = of def]		{GK};
				\node[state]	(gks)	[right = of att]	{GK};
				\node[state, fill=red!25, shape=rectangle]	(gc)	[left = of gkc]		{-1};
				\node[state, fill=green!30, shape=rectangle]	(gs)	[right = of gks]	{+1};
				\path[->]

					(gc)	edge[bend right=55]	node			{}			(mid)
					(gkc)	edge[bend right=15]	node			{}			(def)
							edge[]				node			{}			(gc)
					(def)	edge[bend right=15]	node			{}			(mid)
							edge[bend right=10]	node			{}			(gkc)
					(mid)	edge[bend right=15]	node			{}			(att)
							edge[bend right=60]	node			{}			(gkc)
							edge[bend right=10]	node			{}			(def)
					(att)	edge[bend right=15]	node			{}			(gks)
							edge[bend right=70]	node			{}			(gkc)
							edge[bend right=60]	node			{}			(def)
							edge[bend right=10]	node			{}			(mid)
					(gks)	edge[]				node			{}			(gs)
							edge[bend right=80]	node			{}			(gkc)
							edge[bend right=70]	node			{}			(def)
							edge[bend right=60]	node			{}			(mid)
							edge[bend right=10]	node			{}			(att)
					(gs)	edge[bend left=55]	node			{}			(mid)
					;
			\end{tikzpicture}%
		}
		\caption{
			5-state Chainball Environment. %
			Here states are labelled with
			GK (goalkeeper), %
			D (defence), %
			M (midfield), and %
			A (attack). %
		}
		\label{fig:chainball-5}
	\end{subfigure}
	\begin{subfigure}[t]{\linewidth}
		\centering
		\includegraphics[width=0.8\linewidth]{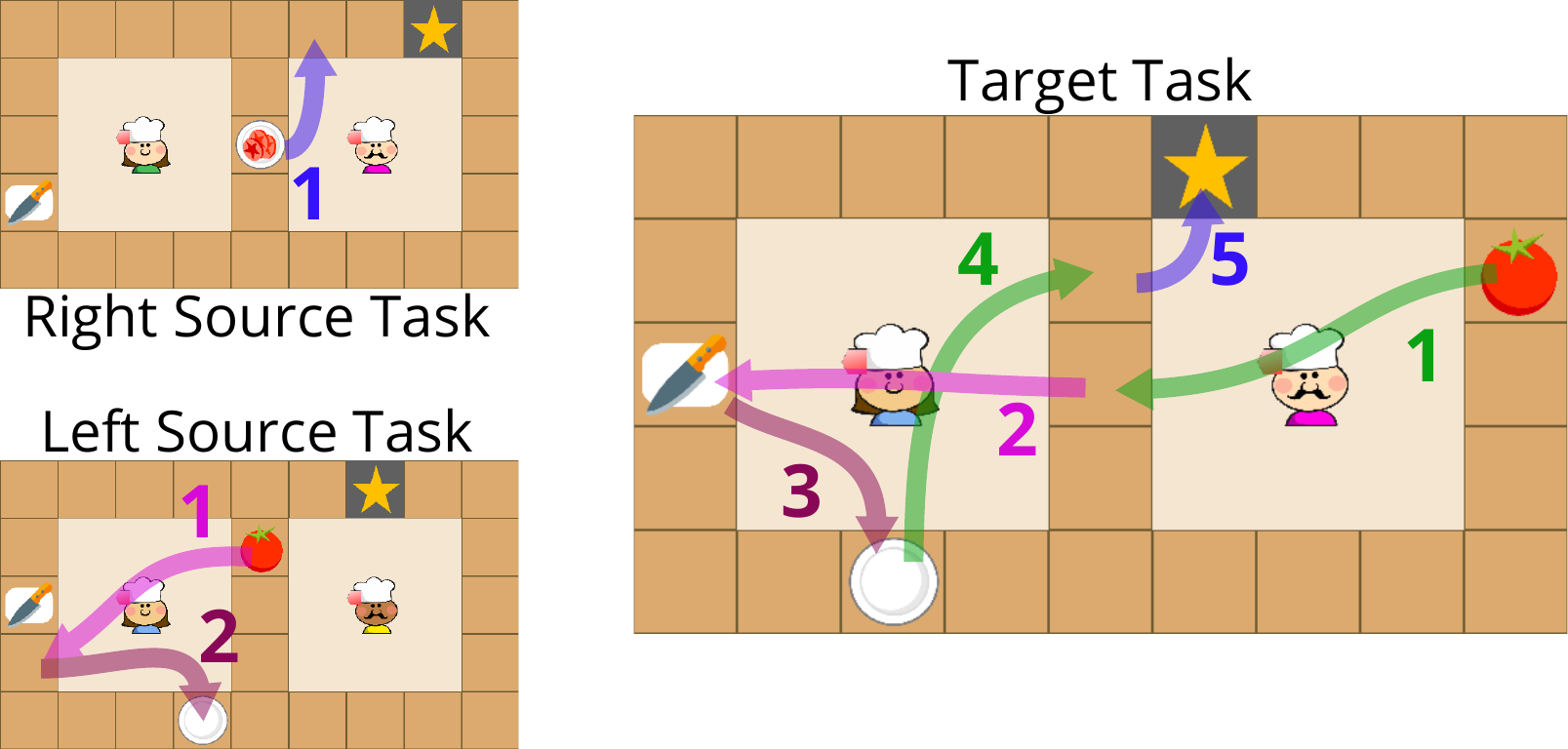}
		\caption{%
		Overcooked sub-task curriculum.
		Arrows represent cooking steps within each task.
		}
		\label{fig:cooking-curric}
	\end{subfigure}
	\begin{subfigure}[t]{\linewidth}
		\centering
		\includegraphics[width=0.7\linewidth]{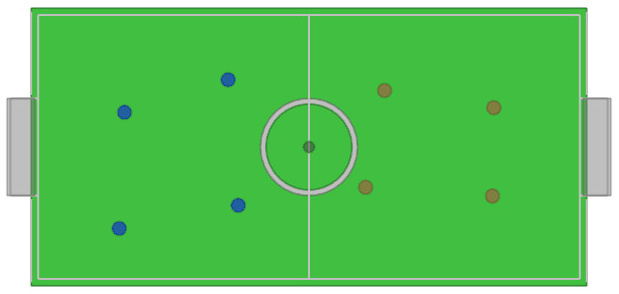}
		\caption{%
			VMAS Football target task. The sub-task curriculum used is similar to that shown in \cref{fig:curric}.
		}
		\label{fig:vmas}
	\end{subfigure}
	\caption{Environments used in our experiments.}
\end{figure}
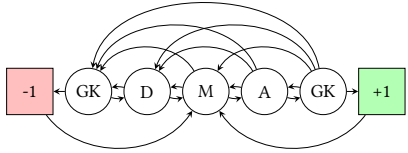
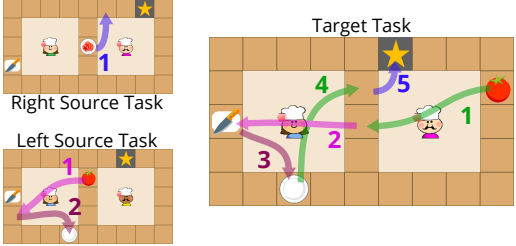
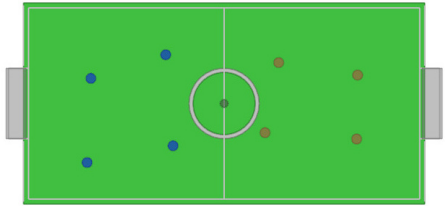

\paragraph{Chainball}\label{sec:chainball}
We introduce the \emph{Chainball} environment as test-bed for the sub-task curriculum approach, mimicking the compositional properties of our football motivating example, while being cheap to evaluate.
Chainball is a difficult task for \ac{MARL} due to the sparse reward and high degree of coordination required.
Chainball (\cref{fig:chainball-5}) is an episodic 4-player game with $|\StateSpace|=11$ discrete states.
The goal of Chainball is to reach the rightmost end of the chain (\enquote{goal scored}), where the team will receive a reward of +1.
However, if the leftmost end of the chain is reached (\enquote{goal conceded}), the team receives a reward of -1.
Upon scoring, the state is reset to the middle state ($\State_6$), and the episode terminates after 90 timesteps.
The transition probability is defined by a matrix for each state --- at timestep $t$, each of four agents chooses an action $\Action_{t,i} \in \qty{1,2,3,4}$, and the probability of moving right is given by the corresponding matrix entry for that state.

Chainball has two 2-player source tasks, Chainball-Att and Chainball-Def, to emulate attack and defence drills respectively.
Each source task consists of $11$ states, but we make $s < 7$ in Chainball-Att and $s > 5$ in Chainball-Def terminal states.
Source task episodes also terminate if a goal is scored or conceded.

\paragraph{Overcooked}\label{sec:overcooked}
We use the \emph{Overcooked} \citep{rotherDisentanglingInteractionUsing2023} environment, which requires multi-step coordination.
The goal of Overcooked is to complete a recipe by moving and processing foods in a grid world.
\Cref{fig:cooking-curric} shows the configuration of our Overcooked sub-task curriculum.
In the target task, agents must coordinate to pass and chop the tomato on the chopping board (1,2),
put the chopped tomato on a plate (3),
and pass the plate with the chopped tomato back to serve at the starred counter (4,5).
The skills to complete steps 2 and 3 can be learned in the \enquote{Right} task;
and skills to complete step 5 can be learned in the \enquote{Left} task.
Steps 1 and 4 require learning new behaviour in the target task.
The team is rewarded for completing each step in the recipe, except steps 1 and 4 in the target task.

\paragraph{VMAS Football}\label{sec:vmas}
The \emph{VMAS Football} environment \cite{bettiniVMASVectorizedMultiagent2022} is a 2D physics-based version of football. VMAS Football has a discrete action space (control inputs in four cardinal directions). Our sub-task curriculum in VMAS football closely follows that shown in \cref{fig:curric}, except we do not use goalkeeper agents. We use a sparse reward of $+1$ for scoring and $-1$ for conceding. Episodes terminate when either team scores, or after 1024 timesteps. The opponent team uses the heuristic policy provided by the VMAS environment.

\begin{figure*}[t]
    \centering
	\begin{subfigure}[t]{0.29\linewidth}
        \includegraphics[width=\hsize]{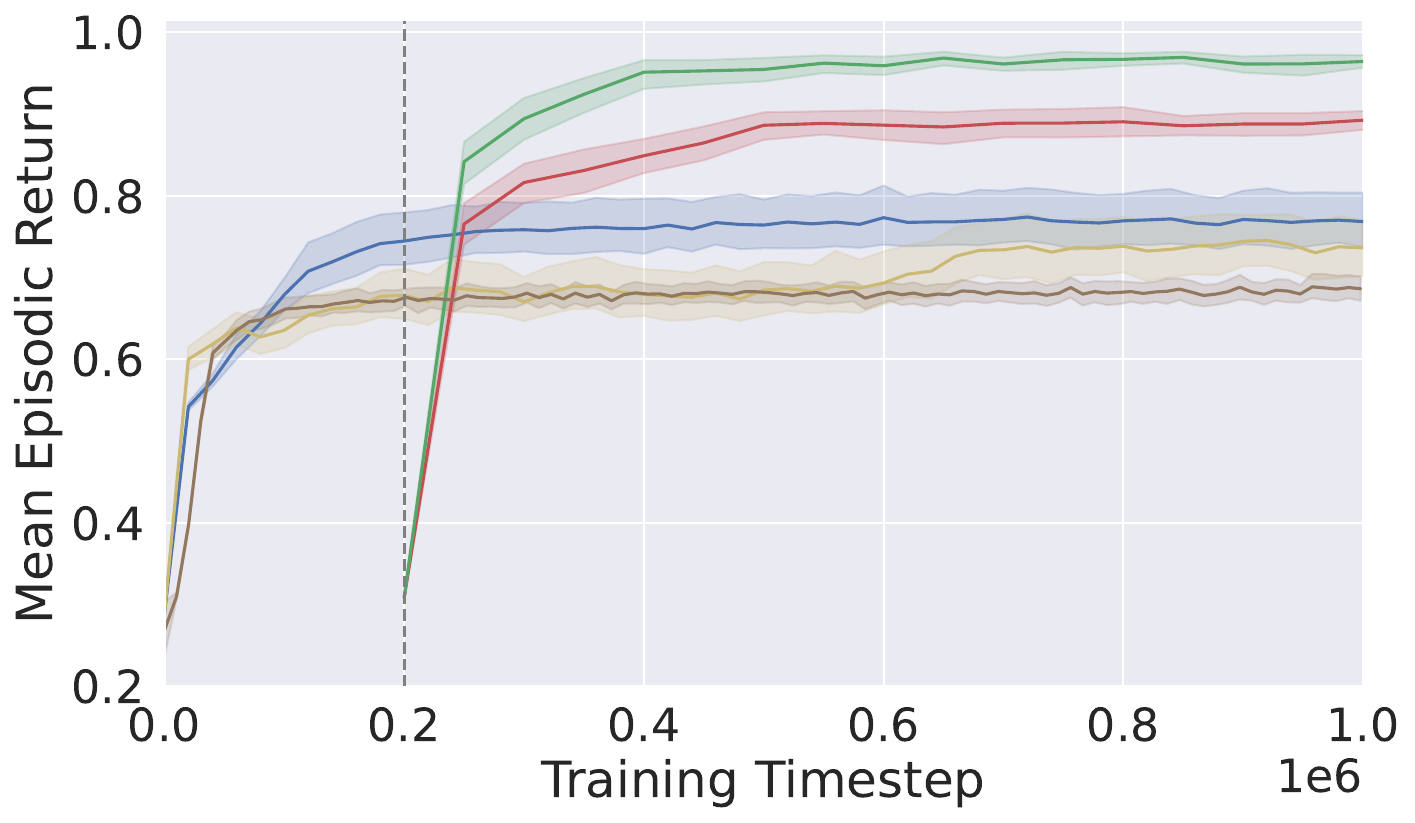}
        \caption{Chainball-11}
        \label{fig:chainball-medoe}
    \end{subfigure}
	\begin{subfigure}[t]{0.29\linewidth}
        \includegraphics[width=\hsize]{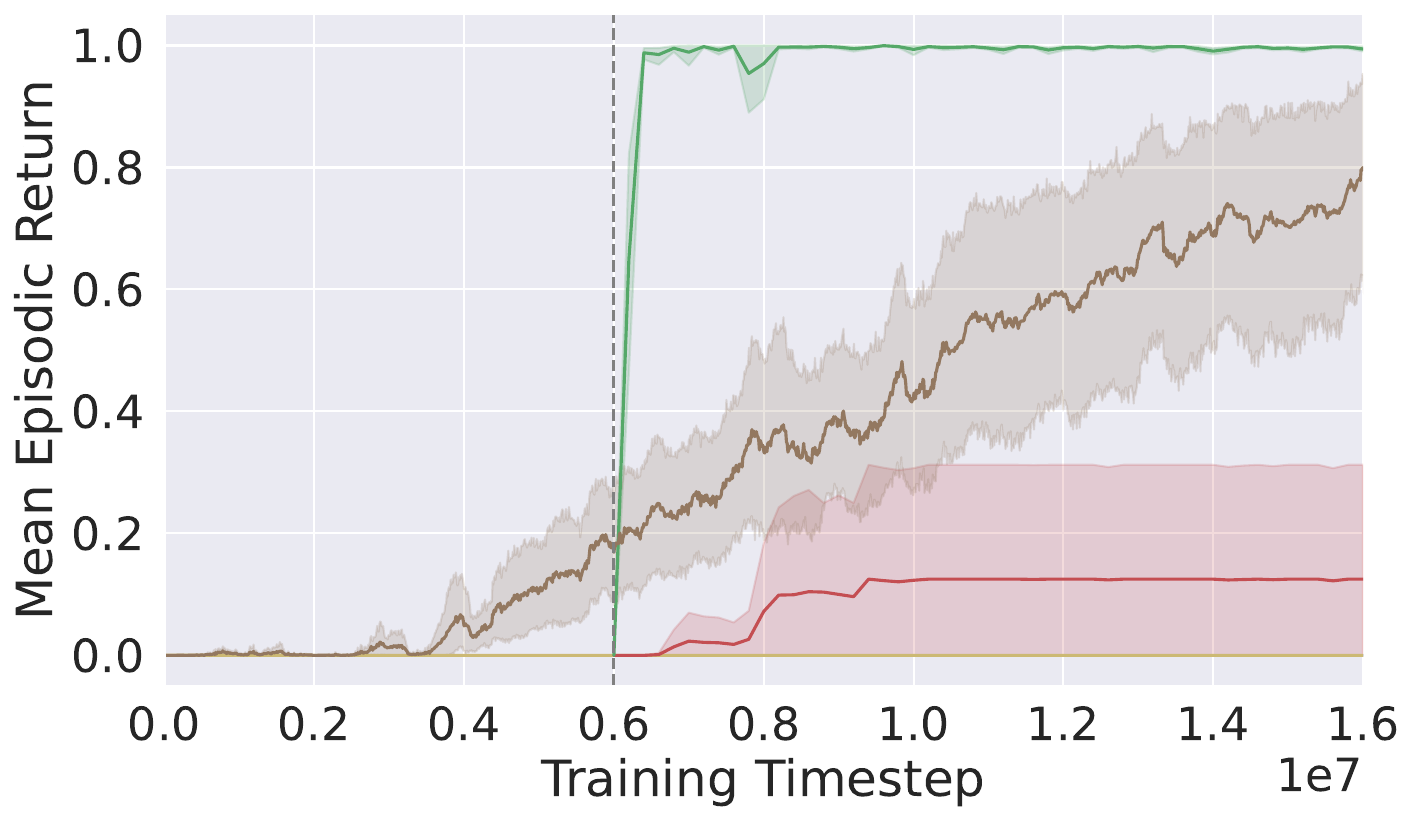}
        \caption{Overcooked}
        \label{fig:cooking-medoe}
    \end{subfigure}
	\begin{subfigure}[t]{0.29\linewidth}
        \includegraphics[width=\hsize]{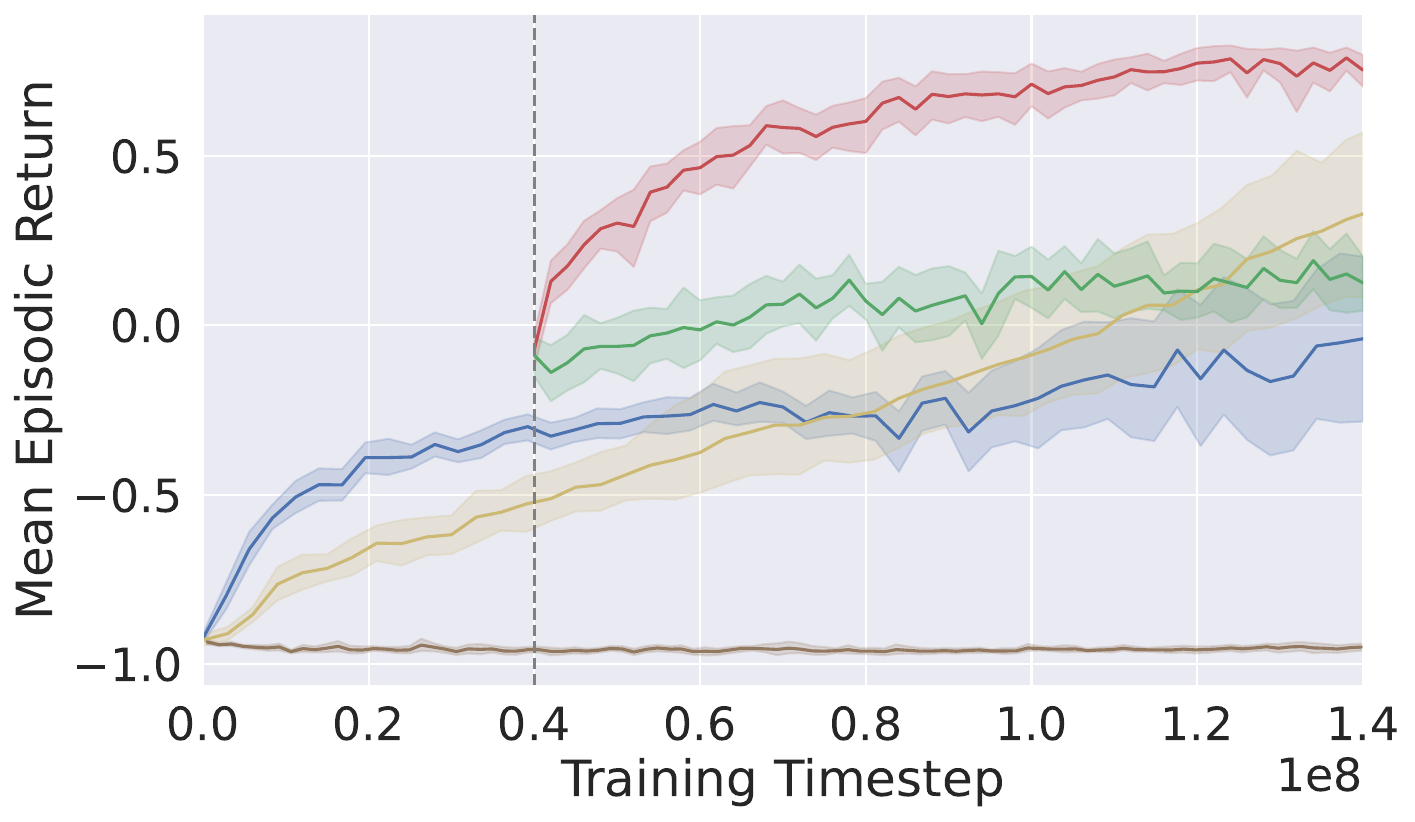}
        \caption{VMAS Football}
        \label{fig:vmas-medoe}
    \end{subfigure}
	\begin{subfigure}[t]{0.11\linewidth}
        \includegraphics[width=\hsize]{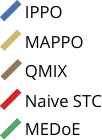}
    \end{subfigure}
    \caption{%
    Target task training returns for each environment.
    Mean episodic returns (100 episodes), averaged over 16 runs (16 different team combinations).
	The \enquote{from-scratch} baselines are averaged over 16 runs (16 seeds; 8 seeds for QMIX on VMAS Football).
    Shaded area shows the 95\% confidence interval of the mean over the runs.
	Naive \ac{STD} and \acs{MEDoE} are shifted on the training step axis to account for the total training steps required across all the sub-tasks, shown by the dashed line.
    }
    \label{fig:medoe-results}
\end{figure*}

\subsection{Protocol and Baselines}\label{sec:baselines}

	In our initial experiments we consider two settings:
	\begin{enumerate*}[label=\arabic*.]
		\item training from scratch in the target task, and
		\item training using a simple sub-task curriculum approach.
	\end{enumerate*}%
	\begin{enumerate}[nolistsep]
		\item For our \enquote{From-Scratch} baselines, we consider the state-of-the-art \ac{MARL} algorithms \acl{MAPPO} (\textbf{MAPPO}) \cite{yuSurprisingEffectivenessPPO2022} and \textbf{QMIX} \cite{rashidQMIXMonotonicValue2018}. We also test the \acl{IPPO} (\textbf{IPPO}) \cite{papoudakisBenchmarkingMultiagentDeep2021} algorithm in a from-scratch setting, as an ablation of our \acl{STD} method which uses IPPO. We run each of these algorithms for 16 seeds directly on the target task.
			Due to computational constraints, in VMAS Football we use 8 seeds.

		\item To investigate the utility of \iacf{STD}, we use a simple approach extending \acs{IPPO}, which we label \textbf{Naive \ac{STD}}.
    For each source task, we generate four seeds of skilled sub-teams by training agents using the standard \acs{IPPO} algorithm (without parameter sharing) until convergence.
	In Chainball, we use tabular actors and critics; whereas in Overcooked and VMAS Football we use neural networks.
    We initialise the target task team networks by cloning final actors and critics from the source tasks.
	This gives 16 final teams, formed by the $4^2$ pairings of source task teams.
	Given these initialisations, Naive \ac{STD} then fine-tunes in the target task using the standard \acs{IPPO} algorithm, and we average the results over these 16 runs.
	For Chainball we use 3 seeds per fine-tuning stage, giving a total of 48 runs.
	Plots in this section also show results for our proposed approach named \textbf{MEDoE}, which we discuss in \cref{sec:medoe}.
	\end{enumerate}
    We describe our hyperparameter tuning protocol and report chosen hyperparameters in \cref{sec:hyperparameter-settings}.

\subsection{Results}\label{sec:results}

Our experiments in \cref{fig:medoe-results} show mixed results for the Naive \acs{STD} approach relative to standard \ac{MARL} baselines. In Chainball, we observe that the use of \iacl{STD} allows the agent to receive a higher converged return than all the from-scratch baselines. Nevertheless, Naive \ac{STD} falls short of achieving the maximum attainable return of $1.0$.
In Overcooked, Naive \ac{STD} outperforms IPPO and MAPPO, which do not learn at all in any of their 16 runs. However, it does not perform as well as QMIX from-scratch. 
In VMAS Football, using a sub-task curriculum enables finding a solution to the task in many fewer timesteps than from-scratch methods.

Contrary to the expectation that the use of a \acl{STD} can make learning a complex target task more efficient, our results show that this is not always the case.
This may explain why, despite the simplicity, naive \acl{STD} approaches have not been widely used or investigated in the \ac{MARL} literature.
We continue to investigate these approaches by asking \emph{why} they sometimes fail to perform as well as hoped. 
Via analysis of our chosen test domains, we conjecture two pitfalls:
coordination difficulties arising from miscoordinated exploration (\cref{sec:coord}); and
issues caused when agents forget useful behaviours they learned during source task training (\cref{sec:forget}).
In \cref{sec:medoe}, we propose an algorithm designed to address these issues, and show that it leads to an improvement in performance in Chainball and Overcooked.

\subsection{Miscoordinated Exploration}\label{sec:coord}
The first problem with naive \aclp{STD} we identify is most evident in the Chainball environment --- miscoordinated exploration, which can lead to convergence upon \emph{Pareto-dominated Nash equilibria} which is a well-known cause of coordination failure in multi-agent learning systems \cite{harsanyiGeneralTheoryEquilibrium1992,clausDynamicsReinforcementLearning1998}.
Pareto-dominated Nash Equilibria in cooperative games are Nash equilibria which are Pareto-dominated by some other Nash equilibrium.
That is, a joint strategy in which no agent has the incentive to \emph{unilaterally} deviate from, but where there exists some other Nash equilibrium with a higher return to \emph{all} players.
It is possible that learning agents can get \enquote{stuck} in Pareto-dominated Nash equilibria if the risks of occasional deviation from the equilibrium strategy are larger in the Pareto-dominated Nash equilibrium than the Pareto-optimal Nash equilibrium.
A classical example of this is the Stag-Hunt game, where independent learning agents -- particularly those which take exploratory actions which cause them to sometimes deviate from the optimal $(\text{Stag},\text{Stag})$ equilibrium -- often converge upon the sub-optimal $(\text{Hare},\text{Hare})$ equilibrium in practice.

Recall that at each timestep, Chainball has a matrix-game structure where agents jointly choose an action which determines the probability with which the players move towards the opponent's goal.
This allows us to perform an equilibrium analysis to demonstrate this issue by examining the policies obtained by fine-tuning agents in the target task using Naive \acs{STD}, as shown in \cref{fig:chainball-medoe}.
Consider the most likely joint action given by the policies at the end of fine-tuning.
For each state, we count the fraction of runs (over our 48 Chainball runs) where the most likely joint action is not the optimal action in that state.
We report results in \cref{tab:chainball-coord}.
Focussing on states 10 and 11 (as the lower states are less frequently visited), we see that convergence to Pareto-dominated Nash equilibria is particularly common in Chainball, and leads to degraded performance of \acl{STD} methods.

\begin{figure}[t]
    \centering
	\includegraphics[width=\hsize]{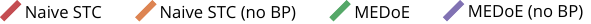}
	\begin{subfigure}[t]{0.48\linewidth}
        \includegraphics[width=\hsize]{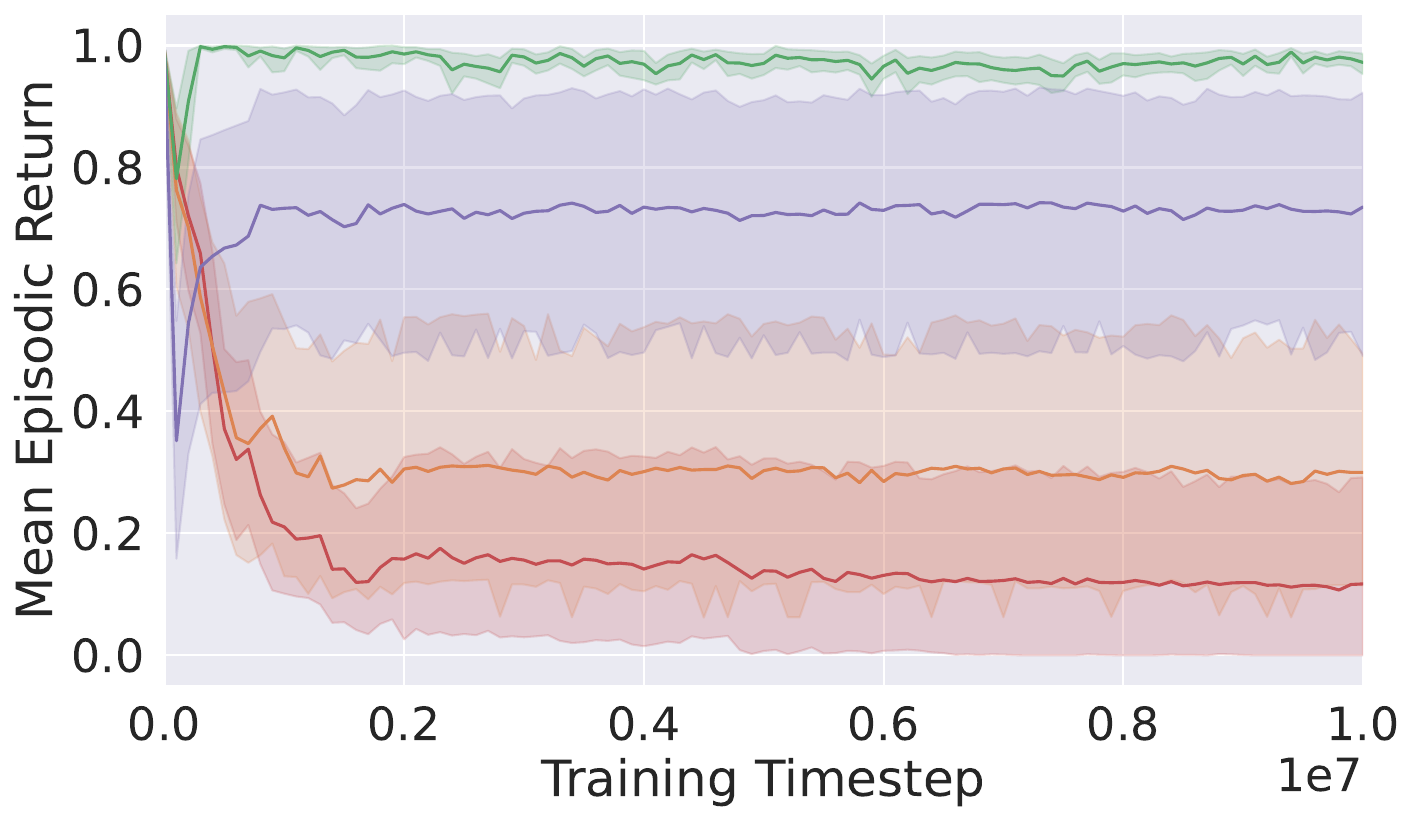}
        \caption{Left Half}
        \label{fig:forget-left}
    \end{subfigure}
	\begin{subfigure}[t]{0.48\linewidth}
        \includegraphics[width=\hsize]{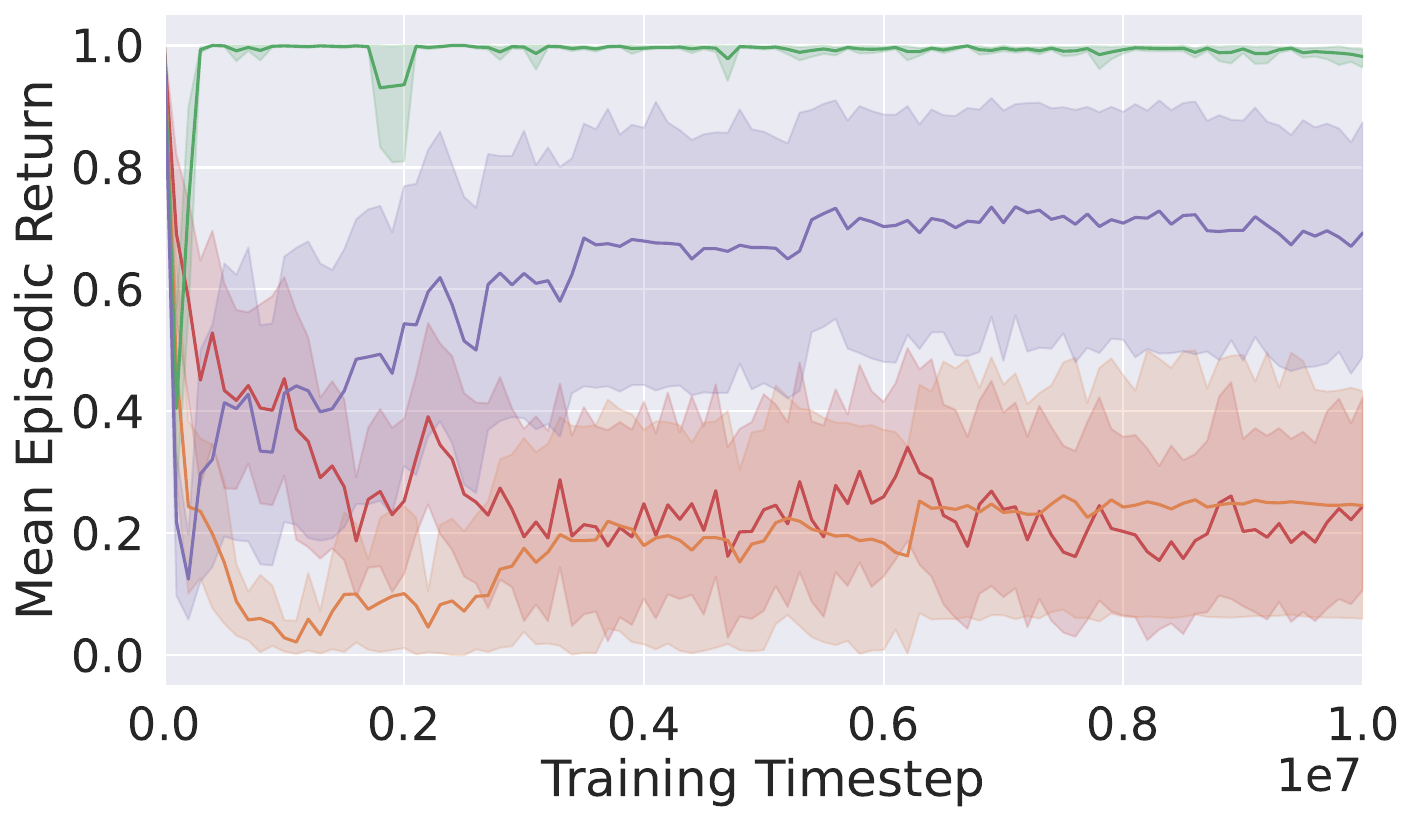}
        \caption{Right Half}
        \label{fig:forget-right}
    \end{subfigure}
	\caption{Forgetting curve in Overcooked.}
\label{fig:forget}
\end{figure}

\begin{table}[]
	\begin{tabular}{@{}p{16mm}rrrrrrrrrrr@{}}%
\toprule
State                                                                   & 1  & 2  & 3  & 4  & 5 & 6 & 7 & 8  & 9  & 10 & 11 \\
\midrule
Sub-optimal  & 90 & 63 & 54 & 38 & 2 & 0 & 2 & 6  & 4  & 23 & 54 \\
PD NE       & 7  & 3  & 38 & 27 & 0 & - & 0 & 0  & 50 & 91 & 58 \\
PD $\epsilon$-NE  & 23 & 3  & 38 & 56 & 0 & - & 0 & 67 & 50 & 91 & 81 \\
\bottomrule
\end{tabular}
\caption{
	Chainball equilibrium analysis for Naive \acs{STD}, showing for each state the percentage of runs where the agents' most likely joint action is sub-optimal;
	and the percentage of those sub-optimal joint-actions which are also Pareto-Dominated Nash Equilibria (PD NE) or $\epsilon$-NE with $\epsilon=0.05$.
}
\label{tab:chainball-coord}
\end{table}

\subsection{Forgetting}\label{sec:forget}

We identify a problem prominent in Overcooked --- agents forgetting useful behaviours.
Consider the Overcooked target task (initial state shown in \cref{fig:cooking-curric}).
The \enquote{right} agent needs to learn step 1. However, until it does so, neither agent receives a reward signal.
As we use entropy regularisation, this means that both agents' policies will gradually reset towards uniform policies until eventually steps 1 and 2 are completed.
We hypothesise that this gradual resetting causes agents to gradually forget skills relevant to steps 2, 3, and 5.

To investigate this hypothesis, we perform an experiment in which we take agents trained in source tasks and fine-tune them in the target task.
As we fine-tune the agents, we periodically re-evaluate the agents \emph{in their respective source tasks} to test whether they have retained the skills learned in their source task.
The results shown in \cref{fig:forget} show evidence of forgetting — as agents are fine-tuned on the target task, their sub-task performance drops, corresponding to a loss of skill on the sub-task.

\section{Modulating Exploration and Training via Domain of Expertise}\label{sec:medoe}
In this section, we introduce a novel approach, \emph{\acf{MEDoE}}, designed to facilitate efficient learning in the fine-tuning stage of a sub-task curriculum.
We focus on addressing the identified problems of
miscoordinated exploration (\cref{sec:coord})
and forgetting (\cref{sec:forget}).
\ac{MEDoE} comprises two key components:
a module which predicts based on source task information whether an agent's policy is likely to be optimal in the current state;
and a module which takes this prediction of expertise to modulate exploration and training hyperparameters.
The key intuitions motivating \ac{MEDoE} are that
\begin{enumerate*}[label=\roman*)]
	\item if an agent's policy is already near-optimal, it should explore less aggressively; and
	\item if an agent's policy is near-optimal then the policy update should be regularised to stay close to this policy in order to prevent forgetting.
\end{enumerate*}
In the following subsections, we discuss each of the two components of \ac{MEDoE} in turn.
We then carry out experiments to show that \ac{MEDoE} can indeed address these issues.

\subsection{Domain of Expertise Classification}\label{sec:classification}
The first component of \ac{MEDoE} is the \acf{DoE} classifier, which for each agent makes a prediction about whether that agent's policy is already close to the optimal decentralised policy for the target task.
\Ac{MEDoE} makes these predictions based on information from the source tasks, using the heuristic that if a target-task state is similar to states observed in an agent's source task, then that agent's policy is more likely to be near-optimal.

We formalise the notion of \ac{DoE} as follows:
Let $\Policy_\iAgent$ be agent $\iAgent$'s current policy.
Let $\PolicySpace^{*,\iTarget}$ be the set of optimal decentralised policies of the target task $\Task_T$.
We consider a target task observation
$\Obs_\iAgent \in \ObsSpace_\iAgent^\iTarget$ to be in the \emph{domain of expertise}, $\DoESet_\iAgent^{\iTarget}$, of agent $\iAgent$ if and only if
\begin{equation}
        \exists \Policy^{*,\iTarget} \in \PolicySpace^{*,\iTarget},\ \KL{\Policy_\iAgent(\cdot|\Obs_\iAgent)}{\Policy^{*,\iTarget}_\iAgent(\cdot|\Obs_\iAgent)} < \tau
    ,\label{eqn:doe-state}
\end{equation}
where $\tau$ is a similarity threshold,
and $D_{KL}$ is the \acs{KL} divergence.

It is impossible to find an algorithm which improves generalisation performance on average across an unconstrained set of source and target tasks \citep{wolpertNoFreeLunch1997}.
We therefore only expect \ac{MEDoE} to work in cases where, at any given time, typically at least one agent's source task policy is optimal in the target task,
i.e.\ $\bigcup_{i\in\Team^\iTarget}{\DoESet_\iAgent^{\iTarget}} \approx \StateSpace^\iTarget$.

Knowing the ground-truth \ac{DoE} set requires knowing the optimal policy, which is not practical.
Instead in \ac{MEDoE} we use experience buffers from each source tasks to infer a \emph{\ac{DoE} Classifier}, ${\DoEC_\iAgent: \ObsSpace_\iAgent^\iTarget \mapsto \bqty{0,1}}$ for each agent $\iAgent$. The ideal \ac{DoE} classifier outputs
\begin{equation}
	\DoEC^*_\iAgent(\Obs_\iAgent) = \begin{cases}
		1 & \text{if } \Obs_\iAgent \in \DoESet_\iAgent^{\iTarget}, \\
		0 & \text{if } \Obs_\iAgent \notin \DoESet_\iAgent^{\iTarget}.
	\end{cases}
\end{equation}
In practice, we formulate the \ac{DoE} classifier training as a binary classification problem where positive examples are taken from agent $\iAgent$'s source task experience buffer, and negative examples are taken from the experience buffers of all agents trained in a different source task to $\iAgent$.
We train \iac{MLP} which learns to identify features of observations which differ between different source tasks, using the heuristic that if the current observation in the target task has features present in a particular source task, then it's likely the agents trained in that source task have expertise.
For example, a classifier trained to distinguish between sample observations from attack drills and defence drills in football might identify the position of the ball on the pitch as a feature of interest.
Defence drills would be identified by the ball being dear our team's goal, which likely corresponds with the \ac{DoE} for defenders.

Concretely, let $\Curric_\iAgent$ represent the source task agent $\iAgent$ was trained in.
Each agent has an associated source task experience buffer ${E_\iAgent = \qty{\Obs_\iAgent^{e}: e\in 1\dots E_\text{max}}}$.
For each agent $\iAgent \in \Team^\iTarget$ we form a dataset 
${\mathcal{D}_\iAgent = \qty{\langle \Obs_j^e, \mathbbm{1}\bqty{\Curric_\iAgent=\Curric_j} \rangle : j\in\Team^\iTarget}}$.
We then train an \ac{MLP} classifier which uses a sigmoid final layer to output a probability $p \in (0,1)$, with the objective of minimising binary cross entropy loss on $\mathcal{D}_\iAgent$.

\subsection{Exploration Modulation}\label{sec:modulation}
The second module of \ac{MEDoE} is the modulation of the exploration and training process during target task fine-tuning, informed by the \ac{DoE} classifier.
In this paper, we focus on our variant of \ac{MEDoE}  based on \ac{PPO} \citep{schulmanProximalPolicyOptimization2017}, which modulates three quantities:
\begin{enumerate}[label=\arabic*), ref=\arabic*]
	\item to address forgetting (\cref{sec:forget}),
		\ac{MEDoE} modulates \begin{enumerate*}[label=\alph*)]
				\item the policy entropy regularisation coefficient, $\EntCoef$ and
				\item the policy behaviour prior coefficient, $\KLCoef$;%
			\end{enumerate*}
	\item to address miscoordinated exploration (\cref{sec:coord}),
		\ac{MEDoE} modulates the policy softmax action selection temperature, $\Temp$.
\end{enumerate}
In the following section, we discuss the intuition behind the modulation of these quantities, and describe \ac{MEDoE}. We provide pseudocode for our algorithm in \cref{sec:algorithm}.

\paragraph{Experts should retain relevant skills}
and non-experts should forget irrelevant skills.
During the source task, agents learn skills which are relevant to the completion of the target task, but also skills which might be irrelevant.
Such irrelevant skills can arise from differences in skill requirements between source and target tasks, or from extrapolation.
Ideally, agents should quickly forget irrelevant behaviours.
However, at the same time, they must retain useful skills, which may be difficult in settings which require complex coordination or with sparse rewards, since forgetting can occur during extended low-reward periods.

To control the rate of forgetting skills, \ac{MEDoE} modulates two parameters.
Firstly, we use entropy-regularised policies, and encourage non-experts to forget irrelevant skills by increasing non-experts' entropy regularisation coefficient, setting ${\EntCoef_\iAgent = \EntCoef_\Base \times \EntBoostCoef^{(1-\DoECL_\iAgent(\Obs_{\iAgent}))}}$,
where $\EntCoef_\Base$ is the base entropy coefficient, and $\EntBoostCoef\geq1$ is the entropy boost coefficient introduced by \ac{MEDoE}.
A high entropy regularisation coefficient increases the rate at which a policy relaxes towards a uniform distribution, intuitively corresponding to an increased rate of forgetting skills.

Secondly, we use fixed behaviour priors \citep{tirumalaBehaviorPriorsEfficient2020} to encourage experts to retain useful skills. This entails using KL-regularised policies (see \cref{eqn:actor_loss}), where we aim to minimise the \ac{KL} divergence between the agent's current policy $\Policy_\iAgent(\Action_\iAgent|\Obs_{\iAgent};\PolicyParams_\iAgent)$, and its frozen source task policy $\Policy_\iAgent(\Action_\iAgent|\Obs_{\iAgent};\bar{\PolicyParams}^\text{BP}_\iAgent)$, thereby encouraging the agent to stay close to its source task behaviour. We boost the KL regularisation coefficient for experts, 
setting ${\KLCoef_\iAgent = \KLCoef_\Base \times \KLBoostCoef^{\DoECL_\iAgent(\Obs_{\iAgent})}}$, where $\KLBoostCoef\geq1$.

\paragraph{Non-experts should explore}
and experts should be predictable to other agents by exploiting existing skills.
By definition, non-expert agents need to learn new behaviours. To do so they must explore.
Exploration in multi-agent systems can have negative effects on learning, such as reducing training stability, and as discussed in \cref{sec:coord}, increasing the difficulty of selecting equilibria which require stable coordination.
We therefore aim to restrict exploration to situations where it is necessary, i.e., when agents are non-experts.
\ac{MEDoE} takes a simple approach: modulate an agent's exploration parameter using that agent's \ac{DoE} classifier.
For the \ac{PPO}-based \ac{MEDoE}, the relevant exploration parameter is the stochastic action selection temperature $T_i$, which scales the policy softmax input:
\begin{equation}
	\Action_\iAgent \sim \Policy(\cdot | \Obs_\iAgent ; \Temp_\iAgent = \Temp_\Base \times \TempBoostCoef^{(1-\DoECL_\iAgent(\Obs_\iAgent))}),\ \forall \iAgent \in \Team,
\end{equation}
where $\TempBoostCoef\geq1$.
During evaluation, we fix the action selection temperature to $\Temp_\Base$. We therefore apply an importance sampling reweighting $\ImpWeight_\iAgent$ (\cref{eqn:imp_weight}) during training:
\begin{equation}
	\ImpWeight_\iAgent = \frac{\Policy_\iAgent(\Action_\iAgent|\Obs_\iAgent;\PolicyParams_\iAgent, \Temp=\Temp_\Base)}{\Policy_\iAgent(\Action_\iAgent|\Obs_\iAgent;\PolicyParams_\iAgent, \Temp=\Temp_\iAgent)}.\label{eqn:imp_weight}
 \end{equation}

Ultimately, we minimise the following policy and value losses for each agent $\iAgent$ in the target team:
\begin{equation}
	\begin{aligned}
	\Loss{(\PolicyParams_\iAgent)} = \label{eqn:actor_loss}%
			& \ImpWeight_\iAgent \text{PPOClip}\pqty{A_\iAgent, \Policy_\iAgent(\Action_\iAgent|\Obs_\iAgent;\PolicyParams_\iAgent), \epsilon} \\%
			& - \EntCoef_\iAgent H(\Policy_\iAgent(\cdot|\Obs_\iAgent;\PolicyParams_\iAgent))  \\
			& + \KLCoef_\iAgent \KL{\Policy_\iAgent(\cdot|\Obs_\iAgent;\PolicyParams_\iAgent)}{\Policy_\iAgent(\cdot|\Obs_\iAgent;\bar{\PolicyParams}^\text{BP}_\iAgent)}, %
	\end{aligned}
\end{equation}
\begin{equation}
	 \Loss{(\VParams_\iAgent)} = \ImpWeight_\iAgent\norm{G_{ t:t+\nStep}-V_\iAgent(\Obs_\iAgent; \VParams_\iAgent)}_2^2 \label{eqn:critic_loss}.
\end{equation}
where $\text{PPOClip}(A,\Policy,\epsilon)$ is the \ac{PPO} policy ratio clipping function described by \citet{schulmanProximalPolicyOptimization2017} with clipping coefficient $\epsilon$.
The advantage function for agent $\iAgent$,
{$A_\iAgent = (G_{ t:t+\nStep}-V_\iAgent(\Obs_{\iAgent,t}; \VParams_\iAgent)),$}
and the $n$-step return for agent $\iAgent$,
$G_{ t:t+\nStep} = \Discount^\nStep V_\iAgent(\Obs_{\iAgent,t+\nStep}; \VParams_\iAgent) + \sum_{i=0}^{\nStep-1}{\Discount^i\Reward_{t+i}}$,
are defined in the usual way.

\subsection{MEDoE Results}\label{sec:medoe-results}

In this section, we present and discuss results from using our proposed approach, \ac{MEDoE}.
We investigate three questions:
\begin{enumerate*}
	\item whether \ac{MEDoE} can improve the performance of using \iacl{STD};
	\item whether behaviour priors alone are responsible for preventing forgetting; and
	\item which of the \ac{MEDoE}-modulated hyperparameters affect performance the most.
\end{enumerate*}

\begin{figure*}
    \centering
	\begin{subfigure}[t]{0.31\linewidth}
        \includegraphics[width=\hsize]{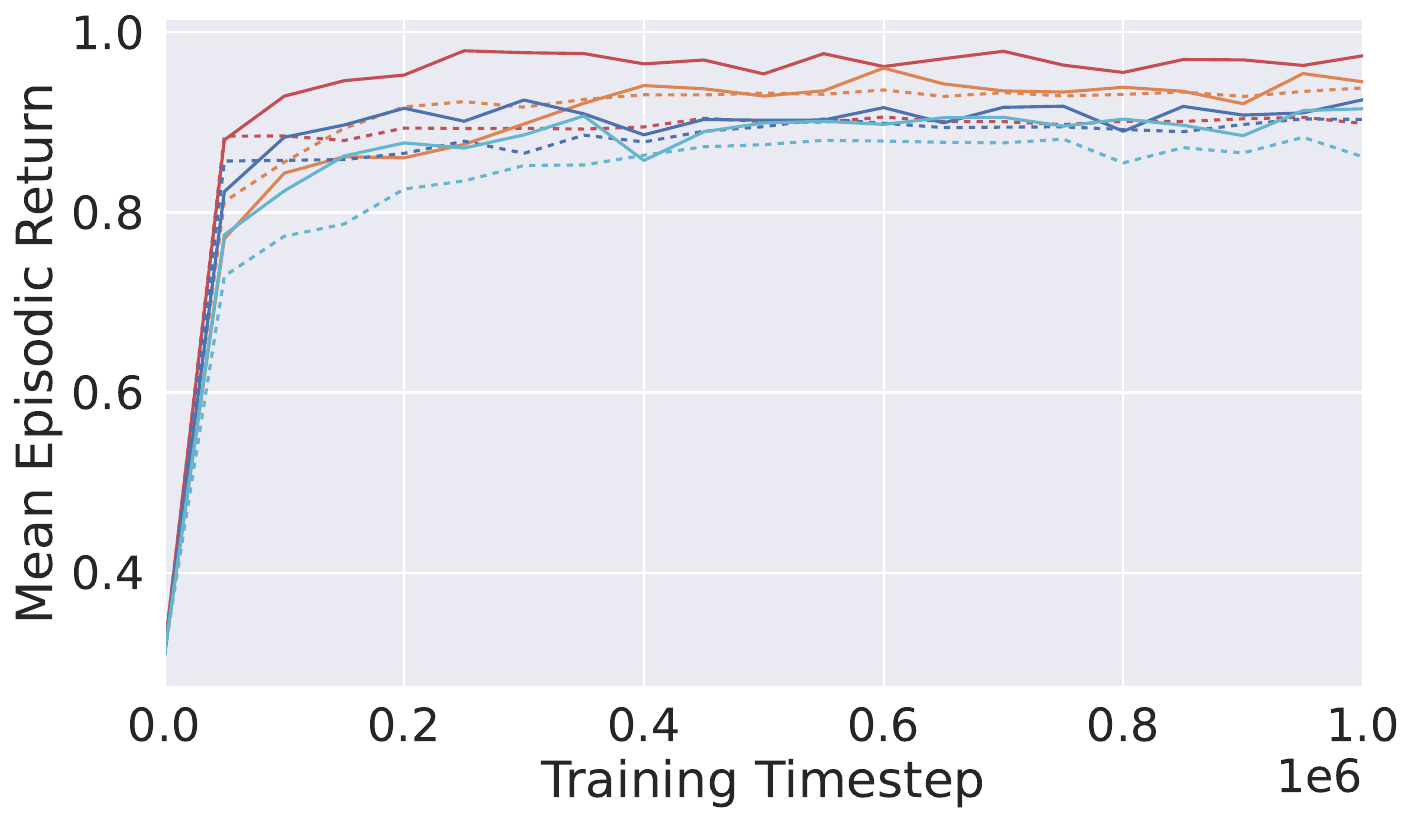}
        \caption{Chainball-11}
        \label{fig:chainball-medoe-ablation}
    \end{subfigure}
	\begin{subfigure}[t]{0.31\linewidth}
        \includegraphics[width=\hsize]{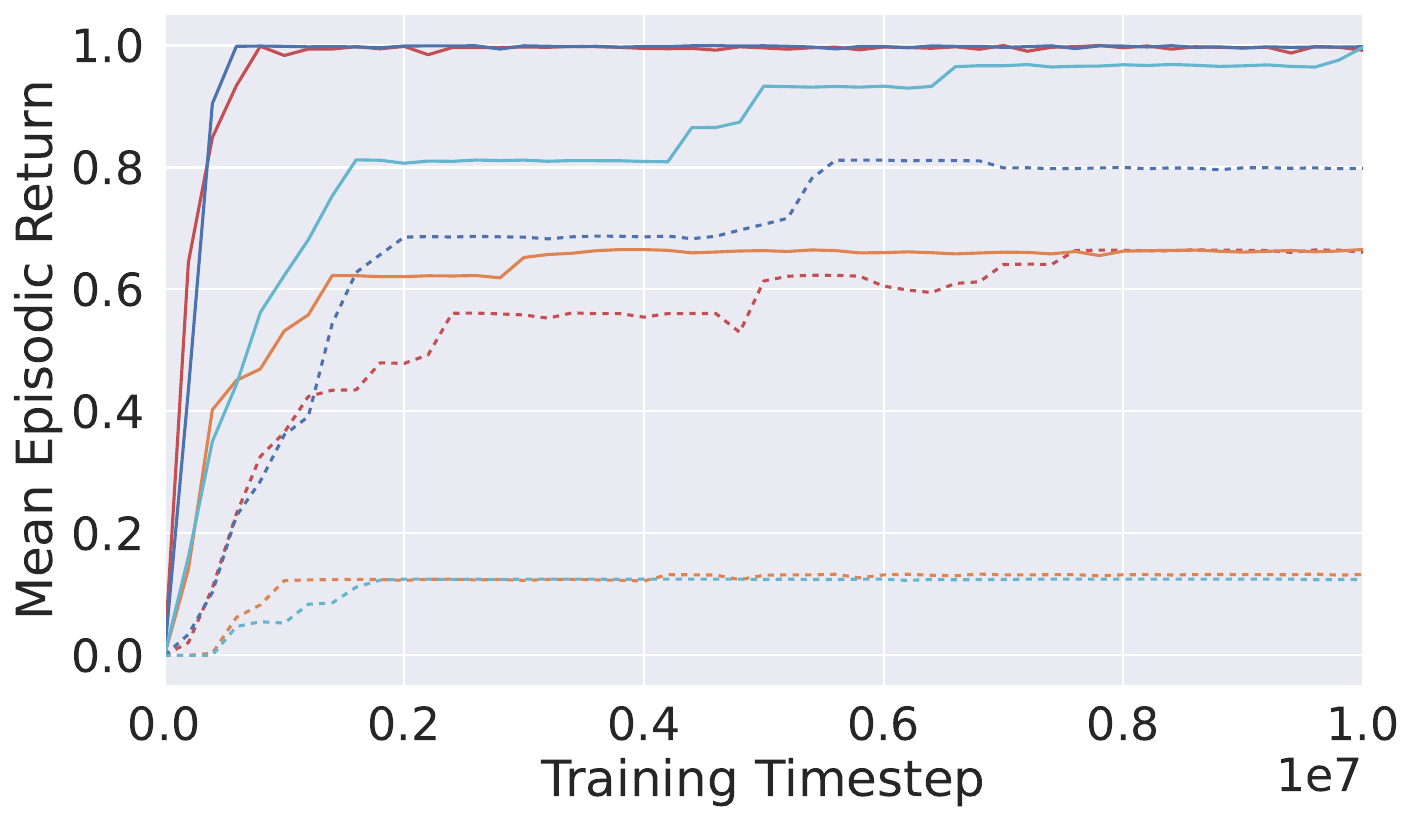}
        \caption{Overcooked}
        \label{fig:cooking-medoe-ablation}
    \end{subfigure}
	\begin{subfigure}[t]{0.31\linewidth}
        \includegraphics[width=\hsize]{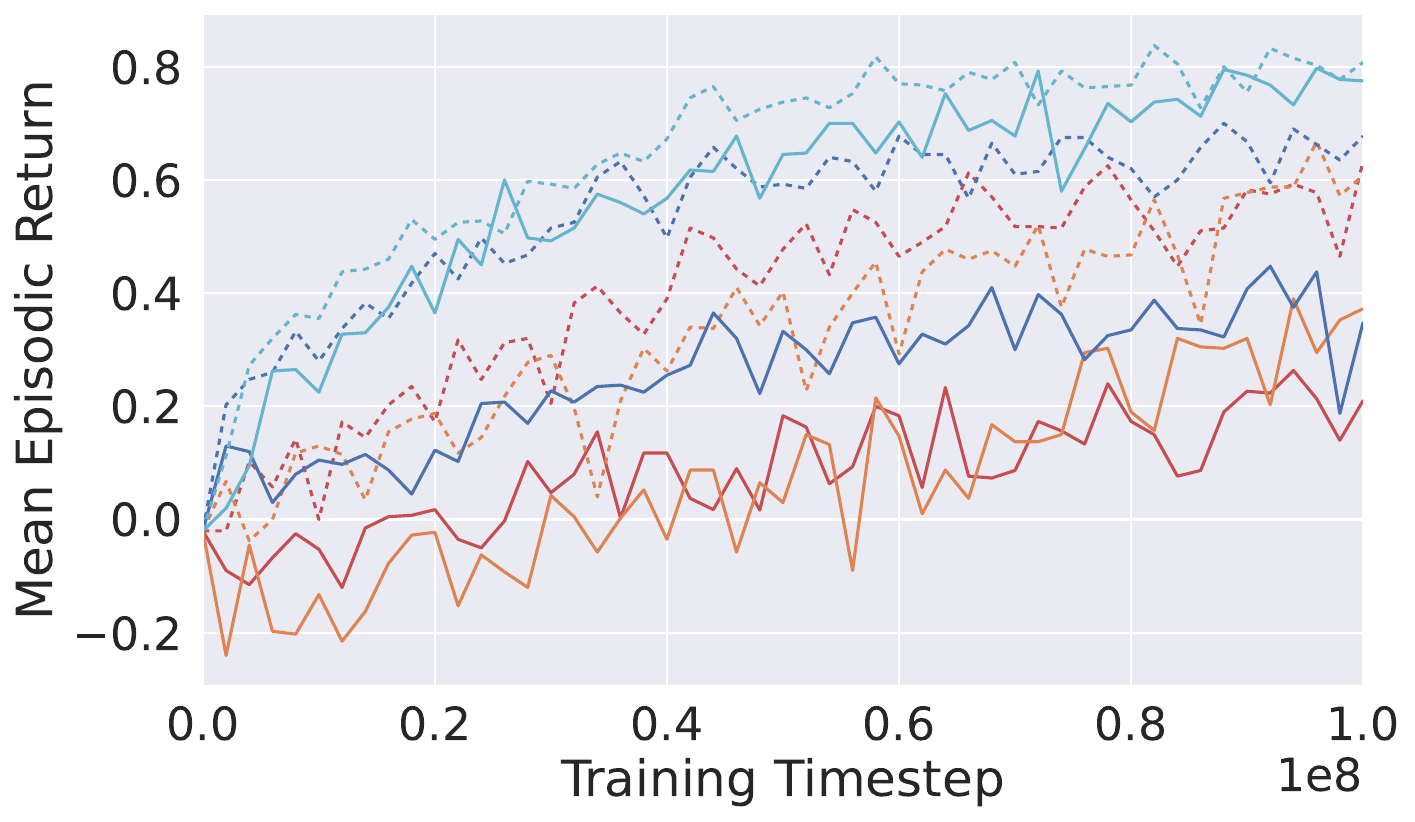}
        \caption{VMAS Football}
        \label{fig:vmas-medoe-ablation}
    \end{subfigure}
	\begin{subfigure}[t]{0.05\linewidth}
        \includegraphics[width=\hsize]{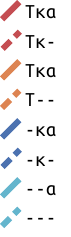}
    \end{subfigure}
    \caption{%
	Ablations of MEDoE.
    Returns are averaged over 16 runs (4 runs in VMAS Football). Error bars are omitted to improve clarity.
	In the legend, labels show a letter if that coefficient is modulated, or a dash otherwise. E.g., \enquote{$T\mbox{--}\alpha$} modulates temperature coefficient ($T$) and entropy coefficient ($\alpha$) but not KL coefficient ($\kappa$).
    }
    \label{fig:medoe-ablation-results}
\end{figure*}

\paragraph{Does MEDoE improve performance?}
In \cref{fig:medoe-results}, we compare our approach, \ac{MEDoE} to the naive sub-task curriculum approach described in \cref{sec:naive-stc}.
We find that for Chainball and Overcooked, \ac{MEDoE} significantly outperforms the from-scratch baselines where the naive sub-task curriculum approach did not, solving the task in many fewer timesteps than all other baselines.
This suggests that \ac{MEDoE} can indeed address the problems identified in \cref{sec:naive-stc}.
However, we also find that \ac{MEDoE} does not improve performance in VMAS Football relative to a naive sub-task curriculum, where the naive approach already significantly outperforms from-scratch baselines.
Our ablation study (\cref{fig:vmas-medoe-ablation}) suggests that temperature modulation is particularly harmful to the performance of \ac{MEDoE} in VMAS Football.
We hypothesise that this is because the quasi-continuous nature of VMAS Football requires effective exploration over temporally extended sequences \cite{osbandDeepExplorationBootstrapped2016}, whereas \ac{MEDoE}'s instantaneous increase in action selection temperature merely leads to poor performance (e.g., \enquote{dithering} causing an inability to consistently move in a straight line).
Future work could therefore consider coupling a \ac{MEDoE}-like approach with hierarchical \acs{RL} to improve performance in (quasi-)continuous control settings.

\paragraph{Are behaviour priors sufficient to address forgetting?}
\Ac{MEDoE} adds a behaviour prior \cite{tirumalaBehaviorPriorsEfficient2020} term to the policy loss to prevent expert agents from forgetting useful skills.
We investigate whether it is the behaviour priors alone which cause the improved performance due to preventing forgetting, or whether \ac{MEDoE} provides benefits of its own. To do so, we present an additional baseline where we augment the naive sub-task curriculum approach with behaviour priors.
\Cref{fig:medoe-results} shows that although the inclusion of behaviour priors alone does lead to increased performance of the naive sub-task curriculum approach, we find that \ac{MEDoE} still outperforms this augmented baseline.
This is particularly evident in Overcooked, which was the motivating case for controlling the rate of forgetting.

Secondly, results from the forgetting experiments (\cref{fig:forget}) show that the use of behaviour priors does indeed slow the rate of forgetting, as expected.
However, we also see that \ac{MEDoE} enables agents to forget less than both from-scratch baselines and the behaviour prior augmented naive sub-task curriculum approach.
Interestingly, with \ac{MEDoE} we see a rapid drop in source task performance during the initial stages of target task fine-tuning, but \ac{MEDoE} then rapidly recovers as the skills are recalled from the behaviour priors.

\paragraph{Which hyperparameters are most important to modulate?}
Our ablation experiments (\cref{fig:medoe-ablation-results}) provide insight into the role the different modulation parameters play in \ac{MEDoE}.
We find that the impact of the different parameters depends on the environment.

In Chainball, we see that each of the hyperparameters contributes somewhat towards \ac{MEDoE}' improved performance. As anticipated, temperature modulation seems to be important to achieving the highest returns in Chainball.
In Overcooked we see that modulating the entropy and behaviour prior KL coefficients is crucial for \ac{MEDoE}'s performance, due to the importance of forgetting in Overcooked.
Modulating the action selection temperature in Overcooked has little consistent effect, perhaps due to the lack of difficult equilibrium selection problems posed by Overcooked compared to Chainball.
In VMAS Football, we see that temperature modulation is actually harmful to the performance of \ac{MEDoE}. This provides evidence for our hypothesis (discussed earlier in this section) that \enquote{exploring} by boosting action selection temperature is a poor strategy in quasi-continuous settings like VMAS Football.
However, we also find that modulating KL and entropy coefficients does not improve performance in VMAS Football.
We believe this is particular to the policies learned in VMAS Football --- we observe that a team can achieve high performance provided one agent learns to skilfully evade the opponent team to score, while the other agents simply defend in the rare cases it is necessary. This reduces the advantage to agents retaining their sub-task skills in VMAS Football.

\section{Related Work}\label{sec:related}

To address challenges in multi-agent reinforcement learning, prior methods also investigated modulating training parameters.
In WoLF-PHC \citep{bowlingRationalConvergentLearning2001} and extensions \citep{bowlingConvergenceNoRegretMultiagent2004}, 
each agent's policy learning rate is modulated according to the intuition that an agent's policy learning rate should be high when it is underperforming relative to its expectations, and low otherwise.
 MA2QL \citep{suMA2QLMinimalistApproach2022} focusses on a team learning setting. 
 MA2QL tackles non-stationarity by allowing only one agent to learn at a time, rather than learning simultaneously.

Work by \citet{vrancxTransferLearningMultiagent2011} considers transfer from simple tasks to complex target tasks in multi-agent systems.
They train a classifier to distinguish between cases in which agents can learn individually, and cases in which they must learn to coordinate as part of a team.
Though similar to our \ac{DoE} classifier, one key difference is that our \ac{DoE} classifier attempts to classify states in which further learning is not required, rather than states in which agents continue to learn without paying attention to other agents.

\citet{wangFewMoreLargescale2020} also accelerates learning of complex multi-agent tasks using a curriculum.
However, that work focusses on cases in which the number of agents is progressively increased throughout the curriculum.
By contrast, our work considers cases in which the task is decomposed based not upon the number of agents, but upon the different skills required by agents in the target task.
Similarly, \citet{tangTransferableMultiagentReinforcement2022} consider scenarios in which agents join an unfamiliar team, and have to rapidly learn to adapt to coordinate with the new team to complete a known task.
In contrast with our work, \citet{tangTransferableMultiagentReinforcement2022} vary only the number of agents between source and target tasks, while the underlying dynamics remain the same.

\citet{taylorParallelTransferLearning2019} consider \emph{parallel transfer learning}, which transfers experiences collected in parallel by separate agents into a target agent, similarly to federated reinforcement learning \citep{qiFederatedReinforcementLearning2021}.
In contrast, our work does not directly transfer skills into individual agents, but instead attempts to accelerate the progress of the \emph{team's} performance.
Another transfer learning work by \citet{yangEfficientTransferLearning2021} provides a framework for transferring skills between agents within a task. In our work, one agent's knowledge and skill might not be useful to another agent --- we instead focus on how each agent should use its own knowledge and skills in the new task.

Several single-agent \acs{RL} methods use a multi-agent approach to curriculum learning
\citep{narvekarCurriculumLearningReinforcement2020,dennisEmergentComplexityZeroshot2020,parker-holderEvolvingCurriculaRegretBased2022}.
\citet{wangSkilledPopulationCurriculum2023} consider similar approaches in a \ac{MARL} setting.
However, these methods focus on the \emph{curriculum design} problem (i.e., generating the series of tasks that form the curriculum) by treating it as a two-player game.
We instead focus on the problem of accelerating learning at each stage in the curriculum, when the target task and decomposed sub-tasks are given as input by an expert.
Our work can therefore be thought of as a \emph{consumer} of sub-task decompositions. Future work could consider uniting \ac{MEDoE} with methods that generate a sub-task decomposition.
To our knowledge, there are few works which consider the challenge of automatic sub-task decomposition.
Approaches using large language models \cite{liSemanticallyAlignedTask2023} may be promising for learning to decompose tasks, using knowledge embedded in human text data.

The problem of effective fine-tuning on new tasks often appears in the continual learning literature,
typically in single-agent settings. 
\Citet{nekoeiContinuousCoordinationRealistic2021} propose using the card game Hanabi as a test-bed for continual learning in multi-agent settings.
\Citet{liuMotorControlTeam2022} engineer an approach to train humanoid agents to play 2-vs-2 football, which, like \ac{MEDoE}, uses behaviour priors \citep{tirumalaBehaviorPriorsEfficient2020}.
However, unlike \ac{MEDoE}, their solution is complex and domain-specific.

Some works consider the assignment of agents into different roles, where agents assigned to the same role employ similar policies \citep{wangROMAMultiAgentReinforcement2020}. In our case, we assume roles do not have to be discovered, and are instead provided implicitly via the given sub-task curriculum. We then focus on efficient use of the sub-task curriculum, rather than training directly in the target task.

Finally, our work is inspired by \Acf{AHT}: the problem of designing single agents capable of coordinating on the fly with previously unseen teammates \citep{mirskySurveyAdHoc2022,stoneAdHocAutonomous2010}.
\Citet{fosongFewshotTeamwork2022} extended the \ac{AHT} problem to the \ac{FST} problem wherein separate \emph{teams} of agents must rapidly learn to complete a new task as part of a unified team.
\citeauthor{fosongFewshotTeamwork2022} provide two framings of the \ac{FST} problem: an \emph{ad hoc teamwork framing} and a \emph{curriculum framing}.
Though our work focuses on the latter framing by investigating methods to accelerate the \ac{MARL} training process, \ac{MEDoE} may also be applicable under the former framing.

\section{Conclusion and Future Work}\label{sec:conclusion}
In this work, we investigated the use of \acf{STD} methods as an approach for accelerating learning of complex \ac{MARL} problems.
We found that given a decomposition of a complex teamwork task into simpler sub-task, straightforwardly applying the standard \acs{IPPO} method to fine-tune agents trained in the sub-tasks can sometimes reduce the number of timesteps required to solve the complex task relative to state-of-the art \ac{MARL} baselines which train from scratch on the complex task.
However, we also showed that counter to expectations, the use of \iac{STD} does not necessarily improve performance over from-scratch \ac{MARL} baselines, and does not necessarily converge to optimal policies.
We identified and investigated two issues which hamper the performance of the naive \ac{STD} approach: miscoordinated exploration and forgetting.

To address these issues, we presented \acf{MEDoE}, which modulates relevant hyperparameters of each agent during target task fine-tuning.
Each agent's modulation is controlled by \iacl{DoE} classifier that provides information about whether the agent's policy is likely to be useful in the target task, given the current observation.
We found that \ac{MEDoE} converges to higher returns in fewer timesteps than the naive \ac{STD} approach in two out of three domains.

\Ac{MEDoE} is a computationally cheap and scalable extension to existing \ac{CTDE} actor-critic methods, with total computational cost increasing linearly with the number of agents in a parallelisable manner.
Though our experiments extend the \acs{IPPO} algorithm, other actor-critic methods can be modified to produce a \ac{MEDoE} version, if desired. 
\ac{MEDoE} does not require additional expert knowledge or engineering input beyond that required for a naive \ac{STD}. 
Though it may not always improve performance in all tasks, \ac{MEDoE} is a promising approach we recommend deploying whenever naive \ac{STD} approaches fail.

Our findings open the door for future work on optimal use of sub-task curricula in \ac{MARL}, and may be extended in multiple directions.
Firstly, we assumed the target task and the sub-task curriculum was provided.
In many cases, this may be a straightforward part of the training setup engineering, though sometimes it may be desirable to automatically propose sub-task decompositions based on task descriptions. This is an open challenge \citep{jeonMASERMultiAgentReinforcement2022}, though recent works employing large language models for this task (e.g.\ \cite{liSemanticallyAlignedTask2023}) may be promising.
Relatedly, future work in the vein of curriculum learning might focus on learning \emph{optimal} sub-task curricula for a given task.
Secondly, in our experiments we focus on one-step curricula, although the framework presented in \cref{sec:curriculum} can represent curricula of arbitrary tree depth. Future work could investigate multi-step curricula, and whether any additional issues arise from their use.
Finally, we present a simple scheme for learning a \ac{DoE} classifier (\cref{sec:classification}).
We show empirically this simple scheme can be sufficient to attain \ac{MEDoE}'s performance benefits. However, future work approaches to obtaining \ac{DoE} classifiers could be investigated and tested.
\ac{DoE} classifiers that can be updated throughout the fine-tuning stage could be investigated.
Furthermore, our experiments are in fully-observable environments, so future work could test \ac{MEDoE} in partially observable environments and develop new approaches to learning \ac{DoE} classifiers under partial observability.

\bibliographystyle{ACM-Reference-Format} 
\bibliography{main}

\appendix

\section{PPO-MEDoE Algorithm} \label{sec:algorithm}
In this section we present our version of \acs{MEDoE} based on the \acf{PPO} algorithm for discrete action spaces. We present a 1-step return version for clarity and conciseness, but the extension to $n$-step is straightforward. \acs{MEDoE} could also be used to extend other actor-critic algorithms in a similar manner. A python implementation of PPO-MEDoE is provided in our public codebase: see \cref{sec:source-code}.

We also note that \ac{MEDoE} could also be used to extend other actor critic methods. Some modification may be required if the action selection mechanism is not softmax action selection. For example, if using mean-variance networks for continuous control \acf{PPO}, the scale of the variance output could be modulated.

\begin{algorithm}[h!]%
	\caption{PPO-MEDoE (1-step)}
	\begin{algorithmic}
		\REQUIRE Team of agents $\Team$ with policies $\qty{\Policy_\iAgent(\Action|\Obs;\PolicyParams_\iAgent) : \iAgent \in \Team}$ and critics $\qty{V_\iAgent(\Obs;\VParams_\iAgent) : \iAgent \in \Team}$
		\REQUIRE Domain of Expertise classifiers, $\qty{\DoECL_i : \iAgent \in \Team}$
		\REQUIRE Hyperparameters
			$\Temp_\Base,$
			$\EntCoef_\Base,$
			$\TempBoostCoef,$
			$\EntBoostCoef$.
		\STATE Set behaviour priors $\bar{\PolicyParams}^\text{BP}_\iAgent$ given initial policy parameters $\PolicyParams_\iAgent$.
		\STATE Observe initial state $\State_0$
		\FOR{$t = 0$ \TO $\Tmax - 1$}
			\STATE Compute boosted exploration coefficient, $\forall \iAgent \in \Team$
				\begin{equation*}
					\Temp_\iAgent = \Temp_\Base \times \TempBoostCoef^{(1-\DoECL_\iAgent(\Obs_{\iAgent,t}))}
				\end{equation*}
			\STATE Sample action $\Action_\iAgent \sim \Policy_i(\cdot | \Obs_{\iAgent,t} ; \Temp=\Temp_i),\ \forall \iAgent \in \Team$
			\STATE Take joint action $\joint{\Action}$, 
			observe next state $\Obs_{jt+1} \sim \StateTransFunc(\cdot|\State_t, \joint{\Action})$
			and receive reward $\Reward_t = \RewardFunc(\State_t, \joint{\Action}_t)$
			\FOR{$\iAgent \in \Team$}
				\STATE Compute importance weight (with stop grad)
					\begin{equation*}
						\ImpWeight_\iAgent = \frac{\Policy_\iAgent(\Action_\iAgent|\Obs_\iAgent;\PolicyParams_\iAgent, \Temp=\Temp_\Base)}{\Policy_\iAgent(\Action_\iAgent|\Obs_\iAgent;\PolicyParams_\iAgent, \Temp=\Temp_\iAgent)}
					\end{equation*}

				\STATE Compute 1-step return $G_{t:{t+1}} = r_t + \Discount V_\iAgent(\Obs_{\iAgent,t+1}; \VParams_\iAgent)$
				\STATE Compute critic loss 
                \begin{equation*}
                    \Loss{(\VParams_\iAgent)} = \ImpWeight_\iAgent\norm{G_{ t:t+1}-V_\iAgent(\Obs_{\iAgent,t}; \VParams_\iAgent)}_2^2
                \end{equation*}

				\STATE Compute advantage $A_i = G_{t:{t+1}} - V_\iAgent(\Obs_{\iAgent,t}; \VParams_\iAgent)$
				\STATE Compute boosted learning coefficients 
					\begin{align*}
						\EntCoef_\iAgent &= \EntCoef_\Base \times \EntBoostCoef^{(1-\DoECL_\iAgent(\Obs_{\iAgent,t}))}  \\
						\KLCoef_\iAgent &= \KLCoef_\Base \times \KLBoostCoef^{\DoECL_\iAgent(\Obs_{\iAgent,t})}
					\end{align*}
				\STATE Compute actor loss
					\begin{align*}
                     	\Loss{(\PolicyParams_\iAgent)} = 
                                 &\ImpWeight_\iAgent \text{PPOClip}\pqty{A_\iAgent, \Policy_\iAgent(\Action_\iAgent|\Obs_\iAgent;\PolicyParams_\iAgent), \epsilon} \\%
                                 &- \EntCoef_\iAgent H(\Policy_\iAgent(\cdot|\Obs_\iAgent;\PolicyParams_\iAgent))  \\
                                 &+ \KLCoef_\iAgent \KL{\Policy_\iAgent(\cdot|\Obs_\iAgent;\PolicyParams_\iAgent)}{\Policy_\iAgent(\cdot|\Obs_\iAgent;\bar{\PolicyParams}^\text{BP}_\iAgent)} %
					\end{align*}

				\STATE Update $\PolicyParams_\iAgent$ and $\VParams_\iAgent$ using gradient descent
			\ENDFOR
		\ENDFOR
		\RETURN Optimised target task policies $\qty{\Policy_\iAgent(\Action|\Obs;\PolicyParams_\iAgent) : \iAgent \in \Team}$ and critics $\qty{V_\iAgent(\Obs;\VParams_\iAgent) : \iAgent \in \Team}$

	\end{algorithmic}
	\label{alg:ppo-medoe}
\end{algorithm}%

\section{Hyperparameter Settings for Experimental Results} \label{sec:hyperparameter-settings}

\Cref{tab:mappo-params,tab:qmix-params,tab:ippo-params} report the hyperparameters used in our experiments. We describe the hyperparameter tuning protocol in this section. Note that none of our experiments use parameter sharing between agents. Configuration files can be found in the public codebase (see \cref{sec:source-code}).

\subsection{Baselines: MAPPO and QMIX}
MAPPO and QMIX are our state-of-the-art \ac{MARL} baselines. For most hyperparameters, we choose common values derived \cite{yuSurprisingEffectivenessPPO2022,papoudakisBenchmarkingMultiagentDeep2021}.
We tune the remaining hyperparameters via a grid search (QMIZ) random search (MAPPO) over the ranges (log-uniform) and choices (uniform) given in \cref{tab:tuning-baselines}.
We run each randomised setting for one seed in the target task, and compute the average final returns and the area-under-curve (AUC) of the training curve.
We examine by hand the relationship between the hyperparameters and the final returns and AUC, and use this to select tuned hyperparameters, reported in \cref{tab:mappo-params,tab:qmix-params}.
Due to computational constraints, we were unable to perform hyperparameter tuning for VMAS, so chose reasonable hyperparameters based on experience and other works.

\begin{table*}[]
\begin{tabular}{llll}
\toprule
					  &     \textbf{Chainball}               &     \textbf{Cooking}                 &     \textbf{VMAS}                    \\
\midrule
\textbf{MAPPO}        &                                      &                                      &                                      \\
Number of Samples     &           256                        &            64                        &         32                           \\
Entropy Coefficient   & $\text{range}\qty(10^{-5}, 10^{-2})$ & $\text{range}\qty(10^{-5}, 10^{-2})$ & $\text{range}\qty(3\times10^{-5}, 10^{-2})$ \\
Actor Learning Rate   & $\text{range}\qty(10^{-4}, 10^{ 0})$ & $\text{range}\qty(10^{-5}, 10^{-2})$ & $\text{range}\qty(10^{-5}, 10^{-3})$ \\
Critic Learning Rate  & $=2\times\text{Actor Learning Rate}$ & $=2\times\text{Actor Learning Rate}$ & $=2\times\text{Actor Learning Rate}$ \\
Gamma                 & $\qty{0.99, 0.999}$                  &  $\qty{0.99, 0.999}$                 &     0.999                            \\
PPO Epochs            & $\qty{4,8,16,32}$                    &  $\qty{2,4,8,16}$                    &     8                                \\%
\midrule
\textbf{QMIX}         &                                      &                                      &                                      \\
Number of Samples     &           256                        &            64                        &         0         \\
Learning Rate         & $\qty{1,2,4,8} \times \qty{10^{-5},10^{-4},10^{-3},10^{-2}} $ & $\qty{\num{1e-4}, \num{3e-4},\num{1e-3},\num{3e-3}}$ & \num{3e-4}                           \\
Target Update         & $\qty{0.01, 200}$                    & $\qty{0.01, 200}$                    & 0.01                                 \\
Epsilon Anneal Period & $\qty{10000, 30000, 100000, 300000}$ & $\qty{10000, 100000, 1000000, 10000000}$ & 14,000,000                           \\
Gamma                 & $\qty{0.99, 0.999}$                  &  $\qty{0.99, 0.999}$                 &     0.999                            \\
\bottomrule
\end{tabular}
\caption{Tuning information for MAPPO and QMIX baselines}
\label{tab:tuning-baselines}
\end{table*}

\begin{table*}[h]
\begin{tabular}{lp{30mm}p{30mm}p{30mm}}
\toprule
\textbf{Hyperparameter} & \textbf{Chainball} & \textbf{Overcooked} & \textbf{VMAS} \\
\midrule
Gamma                           & 0.999 & 0.99 & 0.999\\
Optimiser                       & Adam            & Adam                                 & Adam                                   \\
Adam $\epsilon$                 & \num{1e-5}      & \num{1e-5}                           & \num{1e-5}                             \\
Reward Standardisation          & True & True & True \\
Network architecture            & FC, ReLU,\newline hidden: [256, 256] & FC, ReLU,\newline hidden: [256, 256] & FC, ReLU,\newline hidden: [256,256]\\
Learning Rate                   & \num{1e-2} & \num{3e-4} & \num{3e-4} \\
Buffer Size                     & 5,000 & 5,000 & 5,000 \\
Hypernet Embed Dimension        & 64 & 64 & 64 \\
Target Update                   & 0.01 & 0.01 & 0.01  \\
Evaluation epsilon              & 0 & 0 & 0 \\
Epsilon Anneal Period           & 30,000,000 & 1,000,000 & 14,000,000 \\
\bottomrule
\end{tabular}
\caption{Final MAPPO parameters}
\label{tab:mappo-params}
\end{table*}

\begin{table*}[h]
\begin{tabular}{lp{30mm}p{30mm}p{30mm}}
\toprule
\textbf{Hyperparameter} & \textbf{Chainball} & \textbf{Overcooked} & \textbf{VMAS} \\
\midrule
Discount rate ($\gamma$)        & 0.99            & 0.99                                 & 0.999                                  \\
GAE $\lambda$                   & 0.95            & 0.95                                 & 0.95                                   \\
PPO Clip Coefficient $\epsilon$ & 0.1             & 0.1                                  & 0.1                                    \\
$n$-steps                       & 4               & 16                                   & 64                                     \\
Optimiser                       & Adam            & Adam                                 & Adam                                   \\
Adam $\epsilon$                 & \num{1e-5}      & \num{1e-5}                           & \num{1e-5}                             \\
Gradient Clipping               & False           & False                                & False                                  \\
Actor learning rate             & 0.004           & 0.0004                               & 0.0004                                 \\
Critic learning rate            & 0.008           & 0.0008                               & 0.0008                                 \\
Entropy coefficient ($\alpha$)  & 0.04            & 0.008                                & 0.0004                                 \\
Actor architecture              & Tabular         & FC, ReLU,\newline hidden: [256, 128] & FC, ReLU,\newline hidden: [256,256,128]\\
Critic architecture             & Tabular         & FC, ReLU,\newline hidden: [256, 128] & FC, ReLU,\newline hidden: [256,256,128]\\
\acs{PPO} epochs                & 32              & 16                                   & 8                                      \\
Parallel environments           & 8               & 32                                   & 256                                    \\
\acs{PPO} num.\ minibatches     & 1               & 1                                    & 1                                      \\
\acs{PPO} value clipping        & False           & False                                & False                                  \\
\bottomrule
\end{tabular}
\caption{Final QMIX parameters}
\label{tab:qmix-params}
\end{table*}

\subsection{IPPO and Sub-task Curriculum}
For our sub-task curriculum approaches, we tune hyperparameters in one of the \emph{source tasks} only, and reuse these hyperparameters in both the other source task for that curriculum, as well as the target task fine-tuning. We also use these hyperparameters for our from-scratch IPPO baseline. We perform a random search over the ranges (log-uniform) and choices (uniform) given in \cref{tab:tuning-source-task}. The \ac{MEDoE}-specific boost coefficient values and behaviour prior KL coefficient are not explicitly tuned --- we use the same values across all environments (\cref{tab:medoe-params}). These values were determined as reasonable values during development of this paper. 

\begin{table*}[]
\begin{tabular}{llll}
\toprule
					 &     \textbf{Chainball}               &     \textbf{Cooking}                 &     \textbf{VMAS}                    \\
\midrule
Tuning Source Task   &        Chainball-Att                 &          Left                        &         Attack Drills                \\
Number of Samples    &           256                        &            64                        &             32                       \\
\midrule
Entropy Coefficient  & $\text{range}\qty(10^{-5}, 10^{-2})$ & $\text{range}\qty(10^{-5}, 10^{-2})$ & $\text{range}\qty(10^{-5}, 10^{-1})$ \\
Actor Learning Rate  & $\text{range}\qty(10^{-4}, 10^{ 0})$ & $\text{range}\qty(10^{-5}, 10^{-2})$ & $\text{range}\qty(10^{-3}, 10^{-1})$ \\
Critic Learning Rate & $=2\times\text{Actor Learning Rate}$ & $=2\times\text{Actor Learning Rate}$ & $=2\times\text{Actor Learning Rate}$ \\
Gamma                & $\qty{0.99, 0.999}$                  &  $\qty{0.99, 0.999}$                 &     0.999                            \\
PPO Epochs           & $\qty{4,8,16,32}$                    &  $\qty{2,4,8,16}$                    &     8                                \\%
\bottomrule
\end{tabular}
\caption{Tuning information for source tasks}
\label{tab:tuning-source-task}
\end{table*}

\begin{table*}[h]
\begin{tabular}{lp{30mm}p{30mm}p{30mm}}
\toprule
\textbf{Hyperparameter} & \textbf{Chainball} & \textbf{Overcooked} & \textbf{VMAS} \\
\midrule
BP KL coefficient ($\KLCoef$)            & 0.005                                      & 0.005                               &  0.005                             \\
Temp.\ boost ($\TempBoostCoef$)          & 4                                          & 4                                   &  4                                 \\
KL coef.\ boost ($\KLBoostCoef$)         & 36                                         & 36                                  &  36                                \\
Ent.\ coef.\ boost ($\EntBoostCoef$)     & 36                                         & 36                                  &  36                                \\
Base temp.\ ($\Temp_\Base$)              & $=1/\sqrt{\TempBoostCoef} = 0.5$           & $=1/\sqrt{\TempBoostCoef} = 0.5$    & $=1/\sqrt{\TempBoostCoef} = 0.5$   \\
Base KL coef.\ ($\KLCoef_\Base$)         & $=\KLCoef/\sqrt{\KLBoostCoef} = $          & $=\KLCoef/\sqrt{\KLBoostCoef} = $   & $=\KLCoef/\sqrt{\KLBoostCoef} = $  \\
Base ent.\ coef.\ ($\EntCoef_\Base$)     & $=\EntCoef/\sqrt{\EntBoostCoef} = $        & $=\EntCoef/\sqrt{\EntBoostCoef} = $ & $=\EntCoef/\sqrt{\EntBoostCoef} = $\\
\bottomrule
\end{tabular}
\caption{MEDoE-specific hyperparameters}
\label{tab:medoe-params}
\end{table*}

\begin{table*}[h]
\begin{tabular}{lp{30mm}p{30mm}p{30mm}}
\toprule
\textbf{Hyperparameter} & \textbf{Chainball} & \textbf{Overcooked} & \textbf{VMAS} \\
\midrule
Discount rate ($\gamma$)        & 0.99            & 0.99                                 & 0.999                                  \\
GAE $\lambda$                   & 0.95            & 0.95                                 & 0.95                                   \\
PPO Clip Coefficient $\epsilon$ & 0.1             & 0.1                                  & 0.1                                    \\
$n$-steps                       & 4               & 16                                   & 64                                    \\
Optimiser                       & Adam            & Adam                                 & Adam                                   \\
Adam $\epsilon$                 & \num{1e-5}      & \num{1e-5}                           & \num{1e-5}                             \\
Gradient Clipping               & False           & False                                & False                                  \\
Actor learning rate             & 0.02            & 0.0004                               & 0.004                                  \\
Critic learning rate            & 0.04            & 0.0008                               & 0.008                                  \\
Entropy coefficient ($\alpha$)  & 0.004           & 0.008                                & 0.004                                  \\
Actor architecture              & Tabular         & FC, ReLU,\newline hidden: [256, 128] & FC, ReLU,\newline hidden: [256,256,128]\\
Critic architecture             & Tabular         & FC, ReLU,\newline hidden: [256, 128] & FC, ReLU,\newline hidden: [256,256,128]\\
\acs{PPO} epochs                & 16              & 16                                   & 8                                      \\
Parallel environments           & 8               & 32                                   & 256                                    \\
\acs{PPO} num.\ minibatches     & 1               & 1                                    & 1                                      \\
\acs{PPO} value clipping        & False           & False                                & False                                  \\
\bottomrule
\end{tabular}
\caption{Final IPPO parameters, used across the IPPO, Naive STC and MEDoE baselines}
\label{tab:ippo-params}
\end{table*}

\section{Additional Environment Details} \label{sec:environment-details}
In this section we provide additional environment details. The implementations and configuration files can be found in our public source code: see \cref{sec:source-code}.
\subsection{Chainball}
\subsubsection{Environment Description}

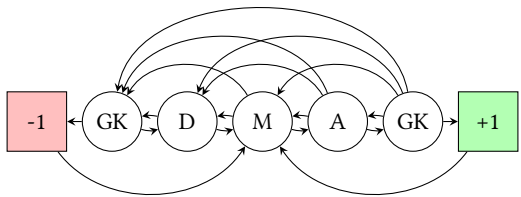
\begin{figure} %
	\centering
	\begin{tikzpicture}%
		[>=stealth,
		 shorten >=0pt,
		 node distance=1cm,
		 on grid,
		 auto,
		]
		\node[state]	(mid)						{M};
		\node[state]	(def)	[left = of mid]		{D};
		\node[state]	(att)	[right = of mid]	{A};
		\node[state]	(gkc)	[left = of def]		{GK};
		\node[state]	(gks)	[right = of att]	{GK};
		\node[state, fill=red!25, shape=rectangle]	(gc)	[left = of gkc]		{-1};
		\node[state, fill=green!30, shape=rectangle]	(gs)	[right = of gks]	{+1};
		\path[->]

			(gc)	edge[bend right=55]	node			{}			(mid)
			(gkc)	edge[bend right=15]	node			{}			(def)
					edge[]				node			{}			(gc)
			(def)	edge[bend right=15]	node			{}			(mid)
					edge[bend right=10]	node			{}			(gkc)
			(mid)	edge[bend right=15]	node			{}			(att)
					edge[bend right=60]	node			{}			(gkc)
					edge[bend right=10]	node			{}			(def)
			(att)	edge[bend right=15]	node			{}			(gks)
					edge[bend right=70]	node			{}			(gkc)
					edge[bend right=60]	node			{}			(def)
					edge[bend right=10]	node			{}			(mid)
			(gks)	edge[]				node			{}			(gs)
					edge[bend right=80]	node			{}			(gkc)
					edge[bend right=70]	node			{}			(def)
					edge[bend right=60]	node			{}			(mid)
					edge[bend right=10]	node			{}			(att)
			(gs)	edge[bend left=55]	node			{}			(mid)
			;
	\end{tikzpicture}
	\caption{
        Chainball-5 Environment. %
        Here states are re-labelled with
        GK (goalkeeper), %
        D (defence), %
        M (midfield), and %
        A (attack). %
    }
    \label{fig:a-chainball-5}
\end{figure}

We introduce the \emph{Chainball} environment as a simple example to test \ac{MEDoE}.
We designed Chainball to mimic the compositional properties of our football motivating example, while allowing for simple evaluation, and use of tabular methods.
\enquote{Chainball-$N$} (\cref{fig:a-chainball-5}) consists of $N$ states, $s \in \qty{1,\dots,N}$.
At timestep $t$, each of four agents chooses an action $\Action_{t,i} \in \qty{1,2,3,4}$.
We define the \emph{forward probability} of taking joint action $\joint{\Action}_t$ in state $s$ as ${f_s(\joint{\Action}_t) = \StateTransFunc(S_{t+1}=s+1|S_t=s, \joint{\Action}_t)}$.
For $s=N$, rather than transitioning to non-existent state $N+1$, the agents score and get a reward of +1, and the state transitions to a restart (kick-off) state (in \cref{fig:a-chainball-5}, the M state).
If the state does not transition forward to state $s+1$, it transitions backwards to state $r < s$ with probability proportional to $1.5^{r-s}$.
This intuitively corresponds to an \enquote{opposing team} getting possession of the ball.
For $s=1$, if the state transition backwards, the team concedes a goal, receiving a reward of -1, and the state transitions to the restart state.
Chainball is an episodic task, which terminates after 90 timesteps.
Chainball has two source tasks, Chainball-$N$-Att and Chainball-$N$-Def, to emulate attack and defence drills respectively.
These source tasks have two agents each.
Each source task consists of $N$ states, but we make states $s < s_\text{Att}$ terminal states in Chainball-$N$-Att, and states $s > s_\text{Def}$ terminal states in Chainball-$N$-Def.
Finally, both source tasks terminate if a terminal state is reached, or if a goal is scored or conceded, or after 90 timesteps.
Our experiments use $N=11$, $s_\text{def} = 6$ and $s_\text{def} = 6$.

For each state, we store a forward probability table $F$, defined such that
\begin{equation}
\begin{aligned}
    F_{s}(a_1,a_2,a_3,a_4) =  \StateTransFunc(S_{t+1}=s+1| & S_t=s, \\
    & \joint{\Action}_t = (a_1,a_2,a_3,a_4)).
    \end{aligned}
\end{equation}
We generate each element of the table uniformly randomly in the interval $[0,0.5]$, and then, for each state we set one of the $4^4=256$ elements to $0.8$ to represent a known optimal joint action. The forward probability table is fixed across all runs. The forward probability table is too large to share in this paper, but can be found in our data disclosure (see \cref{sec:source-code}).

To reduce the difficulty for the full task with 4 players, we make the optimal action in state 5 depend only on the joint action of $(a_1, a_3)$, and the optimal action in state 7 depend only on the joint action of $(a_2, a_4)$:
\begin{align*}
\forall a_1,a_3,a'_1, a'_3, F_5(a_1,a_2,a_3,a_4) &= F_5(a'_1,a_2,a'_3,a_4), \\
\forall a_2,a_4,a'_2, a'_4, F_7(a_1,a_2,a_3,a_4) &= F_7(a_1,a'_2,a_3,a'_4).
\end{align*}

We apply a similar procedure to populate the forward probability tables for the attack and defence source tasks. However, for these source tasks, we do not reduce the difficultly of any states, as each state only has $4^2 = 16$ joint actions.

We design the optimal action in each source task to overlap with the target task.
For example, in Chainball-11-Def, we set the optimal action in states 1,2,3, and 4 for agents 1 and 2 to be equal to the optimal action for agents 1 and 2 in states 1,2,3, and 4 of the target Chainball-11 task.
Similarly, we set the optimal action for agents 1 and 2 in Chainball-11-Att in states 8,9,10,11 to be equal to the optimal action for agents 3 and 4 of the Chainball-11 task.
Outside of these specified states, we require each agent to learn to take actions different to those which were optimal in its source task.

Despite the simplicity of the chainball task, it is difficult to solve due to the sparsity of reward and fact that in most states only one out of 256 joint actions is optimal.

\subsection{Overcooked}
\subsubsection{Environment Description}
\begin{figure}
    \includegraphics[width=\columnwidth]{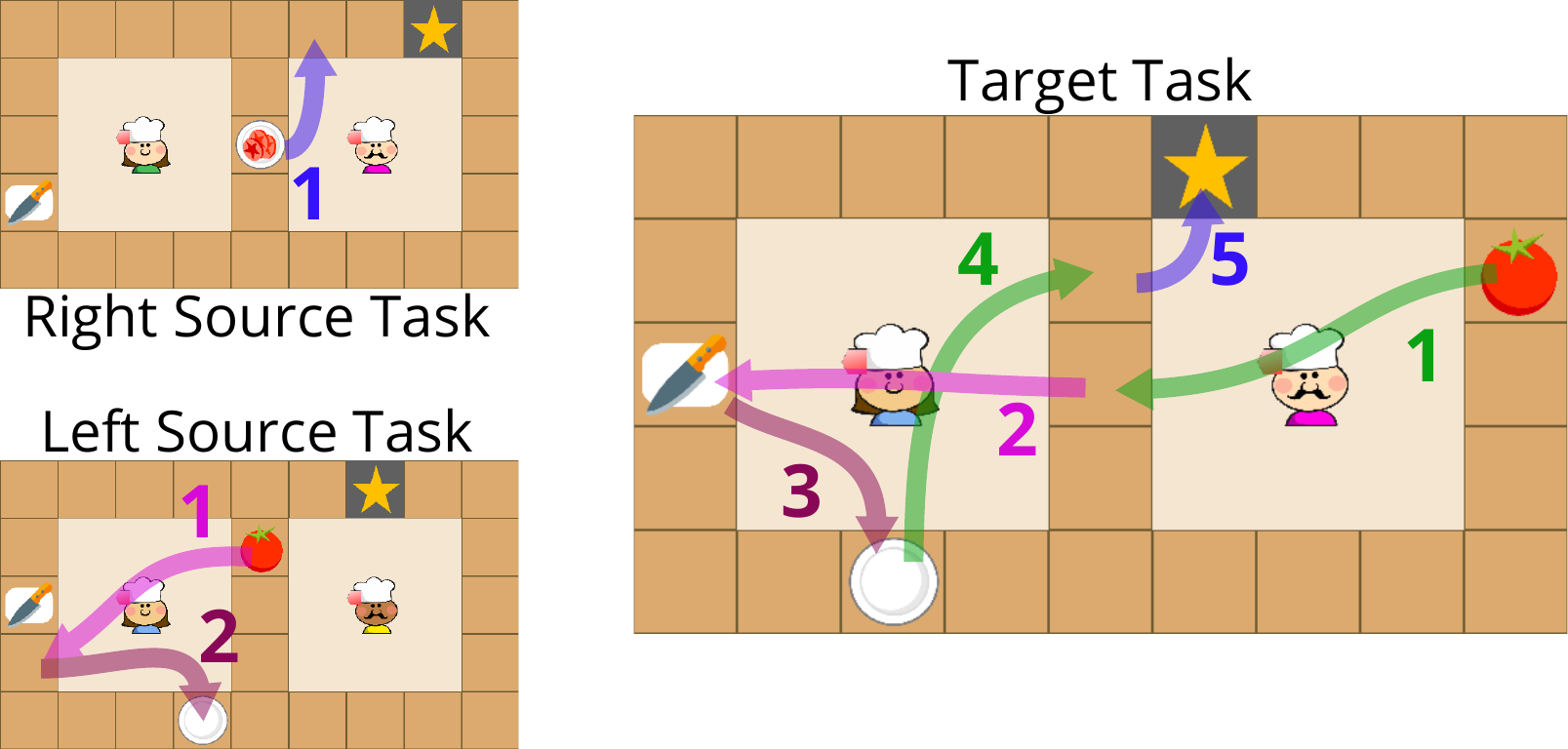}
    \caption{%
    Overcooked Sub-task Curriculum.
    In the target task, agents must coordinate to pass and chop the tomato at the chopping board (1,2),
	put the chopped tomato on a plate (3),
	and pass the plate with chopped tomato back to serve at the starred counter (4,5).
	The skills to complete steps 2 and 3 can be learned in the \enquote{Right} source task;
	and skills to complete step 5 can be learned in the \enquote{Left} source task.
	Steps 1 and 4 require learning new behaviour in the target task.
    }
    \label{fig:a-cooking-curric}
\end{figure}
\emph{Overcooked} is a complex environment which requires multi-step coordination by agents.
The goal of Overcooked is to complete a recipe by moving and processing foods in a grid world.
\Cref{fig:a-cooking-curric} shows the configuration of our Overcooked target task and sub-task curriculum.

The action space in overcooked has 7 actions: 4 cardinal movement actions, a no-op action, and 2 interaction actions: one of which can be used to pick up objects, or put them down on counters; and the other which uses the chopping board when the tomato is placed on it. The object interacted with depends on the orientation of each agent (up/down/left/right).

We use an egocentric observation for each agent which has information about:
\begin{itemize}
\item the ego agent's current absolute location and absolute orientation,
\item the other agent's current relative location and absolute orientation,
\item the relative location and chopped state of the tomato,
\item the relative location of the plate,
\item the relative location of the chopping board,
\item the relative location of the starred delivery tile.
\end{itemize}

We define rewards such that the maximum attainable return in each task is 1. In the full task, the team is rewarded with:
\begin{itemize}
\item +0.267 reward for chopping the tomato on the chopping board,
\item +0.267 reward for putting chopped tomato on the plate,
\item +0.476 reward for delivering the plate to the starred location.
\end{itemize}
In the \enquote{Right} source task, the team is rewarded with +1.0 reward for delivering the plate to the starred location.
In the \enquote{Left} source task, the team is rewarded with +0.5 chopping the tomato, and +0.5 reward for placing the chopped tomato on the plate.

Overcooked terminates once all recipe steps have been completed, or after 100 timesteps.

\begin{figure}
    \includegraphics[width=\columnwidth]{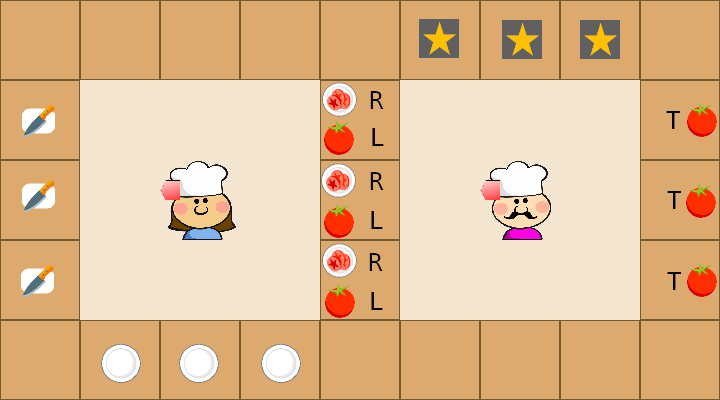}
    \caption{%
    Spawn locations for objects in overcooked.
    Objects are labelled with
    \enquote{L} for \enquote{Left source task spawn position only},
    \enquote{R} for \enquote{Right source task spawn position only}, and
    \enquote{T} for \enquote{Target task spawn position only}.
    }
    \label{fig:a-cooking-spawn}
\end{figure}

\Cref{fig:a-cooking-spawn} shows how we initialise the position of objects in Overcooked.
In each scenario, we uniformly randomly choose a side of the room to spawn the first agent in, placing that agent at the centre of the chosen side; then we spawn the second agent in the centre of the other side.
The other objects are spawned uniformly randomly in:
\begin{itemize}
\item the plate in one of the 3 counter positions on the bottom of the left-hand half of the room,
\item the starred service tile in one of the 3 counter positions on the top of the right-hand half of the room,
\item the chopping board in one of the 3 counter positions on the left of the left-hand half of the room,
\item (Right source task only): the tomato-on-plate on one of the 3 central counter positions,
\item (Left source task only): the tomato on one of the 3 central counter positions,
\item (Target source task only): the tomato on one of the 3 counter positions on the right of the right-hand half of the room.
\end{itemize}

\subsection{VMAS Football}
In VMAS football, we use a pitch which is 2 units long and 1 unit wide. Episodes terminate after a goal is scored or after 1024 timesteps --- whichever is earliest.

\section{Source Code and Data}\label{sec:source-code}
The source code and data associated with this project are available online. See the README in \url{https://github.com/uoe-agents/MEDoE} for more information.


\end{document}